\documentclass[a4paper,twocolumn]{revtex4}  
\usepackage[dvips]{graphicx}
\begin{document}

\title{Noise and aliases in off-axis and phase-shifting holography}

\author{M. Gross and M. Atlan}

\affiliation{
Laboratoire Kastler-Brossel, Laboratoire de Physique de l'\'Ecole
Normale Sup\'erieure, Centre National de la Recherche Scientifique
UMR 8552, Universit\'e Pierre et Marie Curie, 24 rue Lhomond 75231
Paris cedex 05. France}

\author{E. Absil}

\affiliation{
Laboratoire d'Optique, \'Ecole Sup\'erieure de Physique et de Chimie
Industrielles de la Ville de Paris, Centre National de la Recherche
Scientifique UPR A0005, Universit\'e Pierre et Marie Curie, 10 rue
Vauquelin 75231 Paris cedex 05. France}

\date{\today}

\begin{abstract}
We have compared the respective efficiencies of off-axis and
phase-shifting holography in terms of noise and aliases removal. The
comparison is made by analyzing holograms of an USAF target backlit
with laser illumination, recorded with a charge-coupled device
camera. We show that it is essential to remove the LO beam noise,
especially at low illumination levels.\\

\end{abstract}


\maketitle



\section{Introduction}

In digital holography, holograms are recorded by a charge-coupled
device (CCD) array detector, and the image reconstruction is
performed by a computer \cite{Goodmann_1967}, with the major
advantage of avoiding photographic processing.

Off-axis holography \cite{Leith65} is the oldest and the simplest
digital holography technique \cite{Schnars_Juptner_94, Schnars94,
Kreis88}. In off-axis holography, like in holography with
photographic plates \cite{Gabor49}, the reference or local
oscillator (LO) beam is angularly tilted with respect to the object
observation axis. It is then possible to record, with a single
hologram, the two quadratures of the object complex field.
Nevertheless, the object field of view is reduced, since one must
avoid the overlapping of the image with the conjugate image alias
\cite{Cuche00}. Off-axis holography has been applied recently  to
particle \cite{Pu_2004} , polarization \cite{Colomb_2002},  phase
contrast \cite{Cuche99}, synthetic aperture \cite{Massig_2002},
low-coherence \cite{Ansari_2001, Massatsch_2005} and  microscopic
\cite{Massatsch_2005, Marquet_2005} imaging.

In phase-shifting interferometry \cite{Creath1985} and digital
holography \cite{Yamaguchi1997}, one records several images with
different phases of the LO beam to compute the object field in
quadrature in an on-axis (or inline) configuration. Recording a
hologram inline requires a very accurate phase shift between
consecutive images since the conjugate image alias overlaps with
the true image. Aliases are suppressed by making image
substraction. Phase-shifting holography has been applied to 3D
\cite{Zhang_1998, Nomura_2007}, color \cite{Yamaguchi_2002,
Kato_2002}, polarization \cite{Nomura_2007}, synthetic aperture
\cite{Leclerc2001}, low-coherence \cite{Tamano_2006}, surface
shape \cite{Yamaguchi_2006} and microscopic \cite{Zhang_1998,
Guo_2004} imaging.

Recently, we have combined the off-axis configuration
\cite{Schnars_Juptner_94, Schnars94, Kreis88} with our digital
holography  phase-shifting technique \cite{Leclerc2000, Leclerc2001}
to record off-axis phase-shifting digital holograms \cite{gross_07}.
We must notice here that our phase shifting technique minimizes
phase errors \cite{atlan_07}. In the analysis of the data, we have
used a spatial filtering technique \cite{Cuche00} to remove the zero
order and twin image aliases in order to improve the image quality.
By combining all theses techniques, we have perform digital
holography with ultimate sensitivity \cite{atlan_07} i.e. with an
equivalent noise that reach the quantum limit of one photo electron
of noise per reconstructed pixel during the whole measurement time.

In the present paper, we propose to reanalyze the holographic data
used in our previous paper \cite{gross_07} in order to discuss the
respective merits of the different techniques we have combined, i.e.
off-axis   recording of the hologram, spatial filtering, 4-phase
phase-shifting  and low phase error. The discussion will concern
mainly  aliases removal and sensitivity. We have chosen to consider
the data of our previous paper \cite{gross_07} in order to simplify
the discussion, since we have demonstrated in that paper that these
data are sufficient to get optimal sensitivity.


\section{Experimental set-up}

\begin{figure}[]
\begin{center}
\includegraphics[width =8.0cm,keepaspectratio=true]{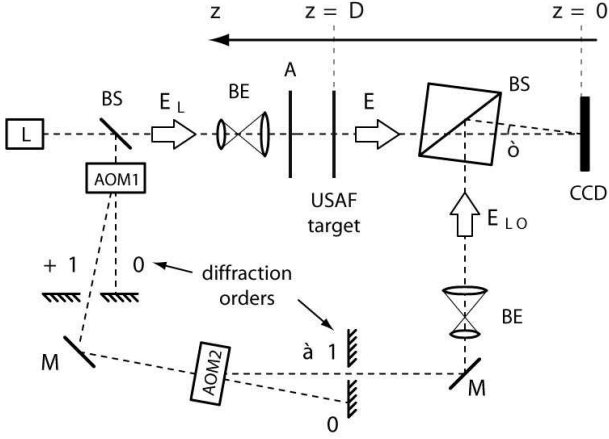}
\caption{ Off-axis, dynamic phase-shifting digital holography setup.
L: main laser; BS: Beam splitter; AOM1 and AOM2: acousto optic
modulators (Bragg cells); BE: beam expander;  M: mirror;  A: light
attenuator. USAF: transmission USAF 1951 resolution target. $E_{L}$,
$E_{LO}$ and $E$ : illumination, local oscillator and object field.
CCD : charge-coupled device array detector.} \label{fig_setup_usaf}
\end{center}
\end{figure}

The holographic setup is sketched in Fig.\ref{fig_setup_usaf}. It
consists of an interferometer in which the beams reaching the
detector are angularly tilted (off-axis configuration) and
frequency-shifted (dynamic phase shift). This setup has been
described in \cite{gross_07}. The main laser L is provided by a
Sanyo DL-7140-201 diode laser ($\lambda=780$ nm, $50$ mW for $95$ mA
of current). It is split into an illumination beam (frequency
$\omega_{L}$, complex field $E_{L}$), and in a reference local
oscillator (LO) beam ($\omega_{LO}$, $E_{LO}$). The object to be
imaged is a back-illuminated USAF target. The optical frequency of
the object field $E$ is the same as the illumination field
($\omega_L$). A set of optical attenuators A (gray neutral filters)
allows to reduce the illumination level. The CCD camera (PCO
Pixelfly digital camera: 12 bit, frame rate $\omega_{CCD}=12.5\, \rm
Hz$, acquisition time $T=125$ ms, with $1280 \times 1024$ pixels of
$6.7 \times 6.7 \mu \rm m ^2$) records the hologram of the object,
i.e. the object ($E$) versus LO ($E_{LO}$) field interference
pattern. By using two acousto-optic modulators (AOM1 and AOM2,
Crystal Technology: $\omega_{AOM1,2} \simeq 80 \, \rm MHz$), the
optical frequency $\omega_{LO}$ of the LO beam, can be freely
adjusted \cite{atlan_07}. To make a 4-phase detection of the object
field $E$, the LO frequency is detuned with respect to the object
field optical frequency by :
\begin{equation}\label{Eq_nu_LO}
    \omega_{L}-\omega_{LO}  = \omega_{AOM2} - \omega_{AOM1} \simeq \omega_{CCD}/4
\end{equation}
Moreover the LO beam is angularly tilted (angle $\theta \sim 1
^\circ$) with respect to the camera-to-object observation axis, so
that the object is seen off-axis with respect to the LO beam
propagation axis. A sequence of 12 CCD images $I_0$ to $I_{11}$
(measurement time $0.96 s$) is recorded, the phase being dynamically
shifted by $\pi/2$ from one image to the next \cite{atlan_07}, as a
consequence of appropriate frequency detuning (Eq.\ref{Eq_nu_LO}).
The interference pattern of the object with the LO field is carried
by the frames $I_m$ recorded by the camera at instants $t_m$.
\begin{eqnarray}\label{equ_t_m}
 t_m = {2\pi m }/{\omega_{CCD}}
\end{eqnarray}

Let us introduce the complex representations ${\cal E}$ and ${\cal
E}_{LO}$ of the signal and LO field enveloppes.
\begin{eqnarray}\label{equ_E_ELO_complex_rep}
   E(t)= {\cal E}e^{j\omega t } + {\cal E}^* e^{-j\omega t }~~~~~~~~\\
   E_{LO}(t)= {\cal E}_{LO}e^{j\omega t } + {\cal E}_{LO}^* e^{-j\omega t }
\end{eqnarray}
where $j^2=-1$. The grabbed CCD signal is proportional to the field
intensity.
\begin{equation}\label{equ_holo2a}
I_m  = \left| {\cal E}~ e^{j\omega t_m } + {\cal E}_{LO}~e^{j\omega_{LO} t_m }  \right|^2\\
\end{equation}
\begin{equation}\label{equ_holo2b}
 I_m = \left| {\cal E}\right|^2+ \left| {\cal E}_{LO}\right|^2+
 {\cal E} {\cal E}_{LO}^*~ e^{j\omega_{CCD}t_m/4}+ c.c.
\end{equation}
where $c.c.$ is the complex conjugate of the ${\cal E} {\cal
E}_{LO}^*$ term. In digital holography, the image is related to the
${\cal E} {\cal E}_{LO}^*~ $ term, while the zero order and twin
images alias are related to the $ \left| {\cal E}_{LO}\right|^2$ and
$c.c.$ terms. The $ \left| {\cal E}\right|^2$ is most often
neglected. Moreover, with the choice of the LO frequency
$\omega_{LO}$, the ${\cal E} {\cal E}_{LO}^*~ $ phase factor becomes
\begin{equation}\label{equ_t_m}
 e^{j\omega_{CCD}t_m/4} \simeq e^{j m\pi/2}=j^{~m}
\end{equation}

\section{Single phase, off-axis holographic images}

In the collected data \{$I_0$, ..., $I_{11}$\}, the LO beam is phase
shifted by $\pi/2$ between consecutive images. To cancel the
phase-shifting effect, we have selected images from the whole set of
12 recorded CCD frames for which the LO phase is the same : $I_0$,
$I_4$ and $I_8$. A single phase, off-axis hologram $H$ in the CCD
plane ($z=0$) is formed as :
\begin{equation}\label{equ_holo}
    H(x,y,0) = I_0(x,y) + I_4(x,y) +I_8(x,y)
\end{equation}

\begin{figure}[]
\begin{center}
\includegraphics[width = 8.2 cm,keepaspectratio=true]{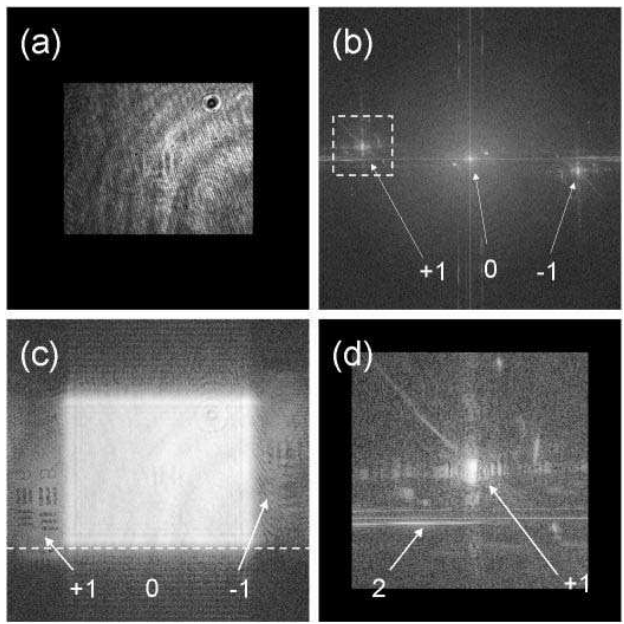}
\caption{Off-axis, single phase reconstruction of the target
image. (a) Hologram $H(x,y,0)$ that corresponds to the $1280
\times 1024$ CCD image padded in a $2048 \times 2048$ zero matrix
(linear gray scale display). (b) k-space field ($2048 \times 2048$
matrix, logarithmic gray scale for the intensity $|\tilde H|^2$).
(c) image calculated on the $2048 \times 2048$ matrix with
logarithmic gray scale display for the intensity $|H|^2$. (d)
Enlargement  of the k-space true image region (dashed rectangle in
Fig.\ref{fig_ima_1ph_700}b). The $400 \times 400$ pixels wide true
image region is copied within a $512\times 512$ zero matrix. The
resulting $512\times 512$ $|H|^2$ intensity image is displayed in
logarithmic scale. (a,d) images correspond to a total signal of
$\simeq 4.3 \times 10^8$ photo electrons for the sequence of 3
images.} \label{fig_ima_1ph_700}
\end{center}
\end{figure}

We have reconstructed the images  by using the standard convolution
method \cite{Schnars94,Kreis2000} that yields  a calculation grid
equal to the pixel size \cite{Zhang2004}. To calculate the
convolution product, we have used the Fourier method, like in
\cite{Leclerc2000}. To avoid reconstruction aliases and to image an
object larger than the CCD size we have enlarged our $1280 \times
1024$ measurement grid  by padding the data into a $2048 \times
2048$ zero matrix \cite{Yamaguchi_2001} (zero padding), as seen on
Fig.\ref{fig_ima_1ph_700}a, where zero is black. Further
calculations are done onto the $2048 \times 2048$ grid.

The reconstructed image is calculated by the following way.
Eq.\ref{equ_holo} yields the real space hologram $H(x,y,0)$ in the
CCD plane. The hologram $\tilde H$ in the CCD reciprocal plane (i.e.
in the $z=0$ k-space) is obtained by discrete Fourier transformation
(FT):
\begin{equation}\label{Eq_image_H(x,0)}
  \tilde H(k_x,k_y,0)= {\rm FT} \left[H(x,y,0)\right]
\end{equation}
The k-space hologram at any distance $z$ from the CCD is then:
\begin{equation}\label{Eq_image_H(k,z)}
  \tilde H(k_x,k_y,z)=  \tilde H(k_x,k_y,0) \, \exp \left( j z ({{k_x}^2 + {k_y}^2
  })/{k}\right)
\end{equation}
where $k=2\pi/\lambda$ is the optical wave vector. The $\exp \left(
j z ({{k_x}^2 + {k_y}^2 })/{k} \right)$ factor is the kernel
function that describe k-space the propagation from $0$ to $z$. The
reconstructed image, which is the hologram in the object plane
($z=D$), is then obtained by reverse Fourier transformation:
\begin{equation}\label{Eq_image_H(x,z)}
  H(x,y,D)={\rm FT}^{-1} \left[\tilde H(k_x,k_y,D)\right]
\end{equation}

Fig.\ref{fig_ima_1ph_700}a  shows the digital hologram $H$ of the
USAF target recorded in the CCD plane. The hologram intensity $|
H|^2$ is displayed in linear gray scale. Raw CCD frames ($1280
\times 1024$ pixels) are padded within a $2048 \times 2048$ zero
matrix displayed in black. Fig.\ref{fig_ima_1ph_700}b shows the
k-space hologram $\tilde H(k_x,k_y,0)$, whose intensity $|\tilde
H|^2$ is displayed in logarithmic scale.

The bright zone (arrow 0) in the center of
Fig.\ref{fig_ima_1ph_700}b corresponds to the FT of the LO beam
($|{\cal E}_{LO}|^2$ term) that is the zero order image. Because
the LO beam is flat in the CCD plane, its Fourier counterpart
yields a very narrow bright spot located in the center of the
k-space, i.e. at $(k_x,k_y) \simeq (0,0)$.

The relevant holographic signal, which corresponds to the ${\cal
E}{\cal E}_{LO}^* $ interference term, is the bright zone of lower
intensity, on the left hand side of the image (arrow +1). Because
the LO beam is off-axis, the FT of the beat signal $ {\cal E}{\cal
E}_{LO}^*$ yields a signal shifted in k-space whose location can
be precisely adjusted by tuning the angle between the LO and
object beams (by tilting the beam splitter in front of the
detector for example).

The twin image, which corresponds to the ${\cal E}^*{\cal E}_{LO}
$ interference term, is symmetrical to the real image with respect
to the k-space center (arrow -1 bright zone in the right hand side
of the image). The angular tilt (represented by $\theta$ on
Fig.\ref{fig_setup_usaf}) between the object and the LO beam
directions defines the separation distance between the true image,
the twin image and the zero order image in k-space.

Fig.\ref{fig_ima_1ph_700}c shows the reconstructed image of the USAF
target whose intensity is displayed in logarithmic scale (i.e.
$|\tilde H(k_x,k_y,z=D|^2$). The true image (arrow +1) is on focus
for a reconstruction distance $z=D=215$ mm in
Eq.\ref{Eq_image_H(k,z)}. The twin image (arrow -1) is blurred. It
would be on focus for the reconstruction distance $z=-D$. The
distance $D$ and the off-axis angle $\theta$ are such that the true
and twin images are partially masked  by the zero order alias, which
is much brighter than the USAF images ($\pm 1$ orders). Nevertheless
because of the logarithmic display the USAF true and twin images are
still visible in Fig.\ref{fig_ima_1ph_700}c.

\begin{figure}[]
\begin{center}
\includegraphics[width = 8.0 cm,keepaspectratio=true]{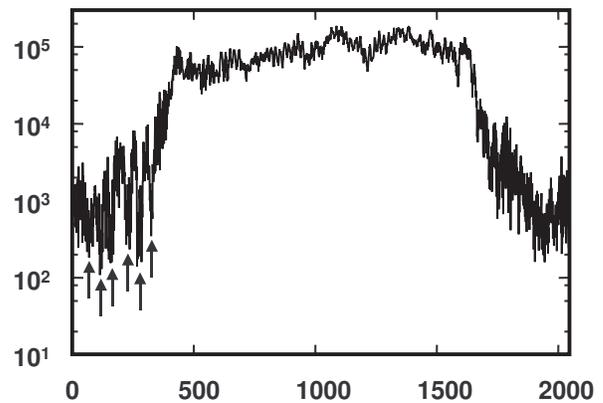}
\caption{Horizontal cut ($y \simeq 1196 $) of the reconstructed
field intensity $|\tilde H|^2$ of Fig.\ref{fig_ima_1ph_700}c.
Vertical axis is $|\tilde H|^2$ in Arbitrary Units (A.U.).
Horizontal axis is pixel index $x=0...2047$. Vertical display is
logarithmic.} \label{fig_cut_1ph_700}
\end{center}
\end{figure}

To perform quantitative study of the Fig.\ref{fig_ima_1ph_700} (c)
image, we have represented an horizontal cut along the
Fig.\ref{fig_ima_1ph_700} (c) dashed line. We have plotted on
Fig.\ref{fig_cut_1ph_700} the reconstructed field intensity
($|\tilde H|^2$) trace at $y \simeq 1196$.  To reduce the noise the
curve is obtained by averaging over 11 pixels ($y=1191$ to
$y=1201$). As seen on Fig.\ref{fig_ima_1ph_700}c, the profile
crosses 6 USAF black vertical bars, which are visible on the left
hand side of the image. The profile crosses also the zero order
image (white rectangle in the center of the image). On
Fig.\ref{fig_cut_1ph_700}, the zero order signal, which correspond
to the central region of the curve ($x=500 $ to $1500$), is much
higher (about $10^5$ A.U.) than the USAF signal (about $5\times
10^3$ at maximum), which is visible on the curve left hand side
($x=0 $ to $400$). The USAF bars (highlighted by arrows) are
nevertheless visible on the curve.

To select the relevant first order image, and to fully suppress the
zero order and twin image aliases, we have used, as proposed by
Cuche et al. \cite{Cuche00}, a k-space filtering (or spatial
filtering) method. We have selected, in the k-space $2048\times
2048$ matrix $|\tilde H(k_x,k_y,z=0|^2$, a $400 \times 400$ region
of interest centered on the true image bright zone (white dashed
rectangle on Fig.\ref{fig_ima_1ph_700}b). Note that this selection
is made possible by the off-axis geometry that has translated the
true image in the left hand side of the k-space domain. The selected
area is then copied in the center of a $512\times 512$ zero matrix
(zero padding) as shown on Fig.\ref{fig_ima_1ph_700}d. The
calculation of the $z=D$ k-space and real space holograms
(Eq.\ref{Eq_image_H(k,z)}) are then done on this $512\times 512$
calculation grid.

\begin{figure}[]
\begin{center}
\includegraphics[width = 8.2 cm,keepaspectratio=true]{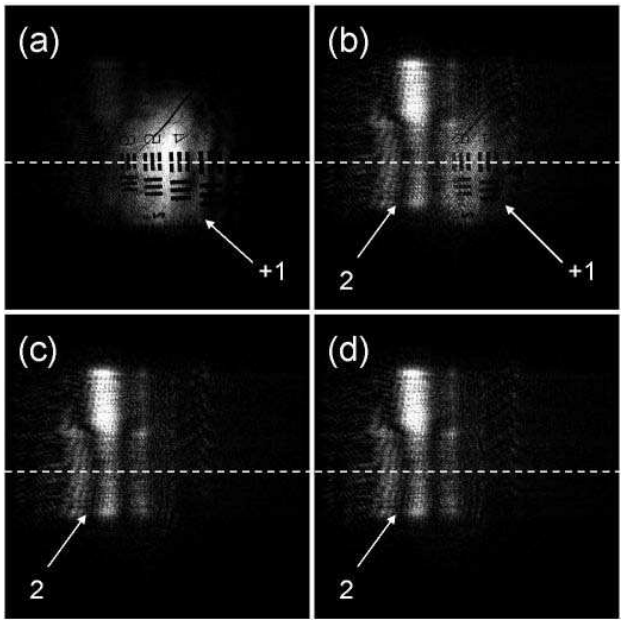}
\caption{Off-axis reconstructed image of a USAF target in
transmission with low light illumination. Images  are obtained with
k-space filtering with $\simeq 4.3\times 10^8$ (a), $8.7\times 10^6$
(b), $8.7 \times 10^4$ (c) and $1.2\times 10^4$ (d) photo electrons
respectively for the sequence of 3 images. $|H|^2$ intensity
displayed in linear gray scale.} \label{fig_usaf_1ph}
\end{center}
\end{figure}

Fig.\ref{fig_usaf_1ph}a shows the object plane real space hologram
$H(x,y,z)$ obtained by computing Eq.\ref{Eq_image_H(x,z)}, which
yields the USAF target image. Note that the translation of the
selected zone in the center of the k-space domain in
Fig.\ref{fig_ima_1ph_700}d moves the reconstructed image of the
USAF target in the center of the image as seen in
Fig.\ref{fig_usaf_1ph}a. The comparison of
Fig.\ref{fig_ima_1ph_700}c with Fig.\ref{fig_usaf_1ph}a
illustrates the ability of the spatial filtering method
\cite{Cuche00} to improve the quality of the reconstructed image
in single phase off-axis digital holography. Nevertheless, we will
see that image quality will degrade when the illumination level
becomes low.

The sensitivity limit of the single phase off-axis configuration
was assessed by recording images of  the USAF target at different
levels of illumination. To get quantitative results, we have
determined the absolute number of photo electrons that corresponds
to the signal beam impinging onto the array detector. The
calibration procedure is described in detail in reference
\cite{gross_07}. Fig.\ref{fig_usaf_1ph} shows the reconstructed
images obtained for various attenuation levels. Although the
Fig.\ref{fig_usaf_1ph} b, c, d images are reconstructed with the
same holograms as in reference \cite{gross_07}, the calibration
factor is slightly different.  In reference \cite{gross_07}, 12
holograms are used ($I_0$ ... $I_{11}$) but the holographic data
are truncated within a $1024 \times 1024 $ calculation matrix.
Here, we make use of 3 CCD frames to form the holograms with the
whole $1280 \times 1024 $ pixel array (cf. Eq.\ref{equ_holo}). The
number of photo electrons is thus lowered by a factor
$1280/(4\times 1024)= 0.31$ in comparison with the results
presented in reference \cite{gross_07}.

On Fig.\ref{fig_usaf_1ph}a, with $\simeq 4.3\times 10^8$ photo
electrons for all the pixels of the set of 3 images (i.e. for
$I_0+I_4+I_8$), one can see the USAF target with a good
$\textrm{SNR}$ (Signal to Noise Ratio). On Fig.\ref{fig_usaf_1ph}b,
with $8.7\times 10^6$ photo electrons, one can see the USAF target
(arrow +1), but a parasitic signal is visible (arrow 2).
%
%
%
When the illumination level goes down, the true image signal (arrow
+1 in Fig.\ref{fig_usaf_1ph}a and b) decreases, while the parasitic
signal (arrow 2 in Fig.\ref{fig_usaf_1ph}b, c and d remains
unchanged. The parasitic signal (arrow 2), which is visible on
Fig.\ref{fig_usaf_1ph}b becomes then dominant on
Fig.\ref{fig_usaf_1ph}c and d with $8.7 \times 10^4$ (c) and
$1.2\times 10^4$ (d) photo electrons, while the USAF target
vanishes.

It is difficult to determine the exact nature of the parasitic
signal. Nevertheless, since the parasites do not depend on the
power of signal beam, they are related to the LO beam alone. We
can thus simply conclude that off-axis holography with spatial
filtering is not sufficient to remove all LO beam parasitic
contributions of our experiment, and is thus unable to reach
optimal detection sensitivity.

\begin{figure}[]
\begin{center}
\includegraphics[width = 4.0 cm,keepaspectratio=true]{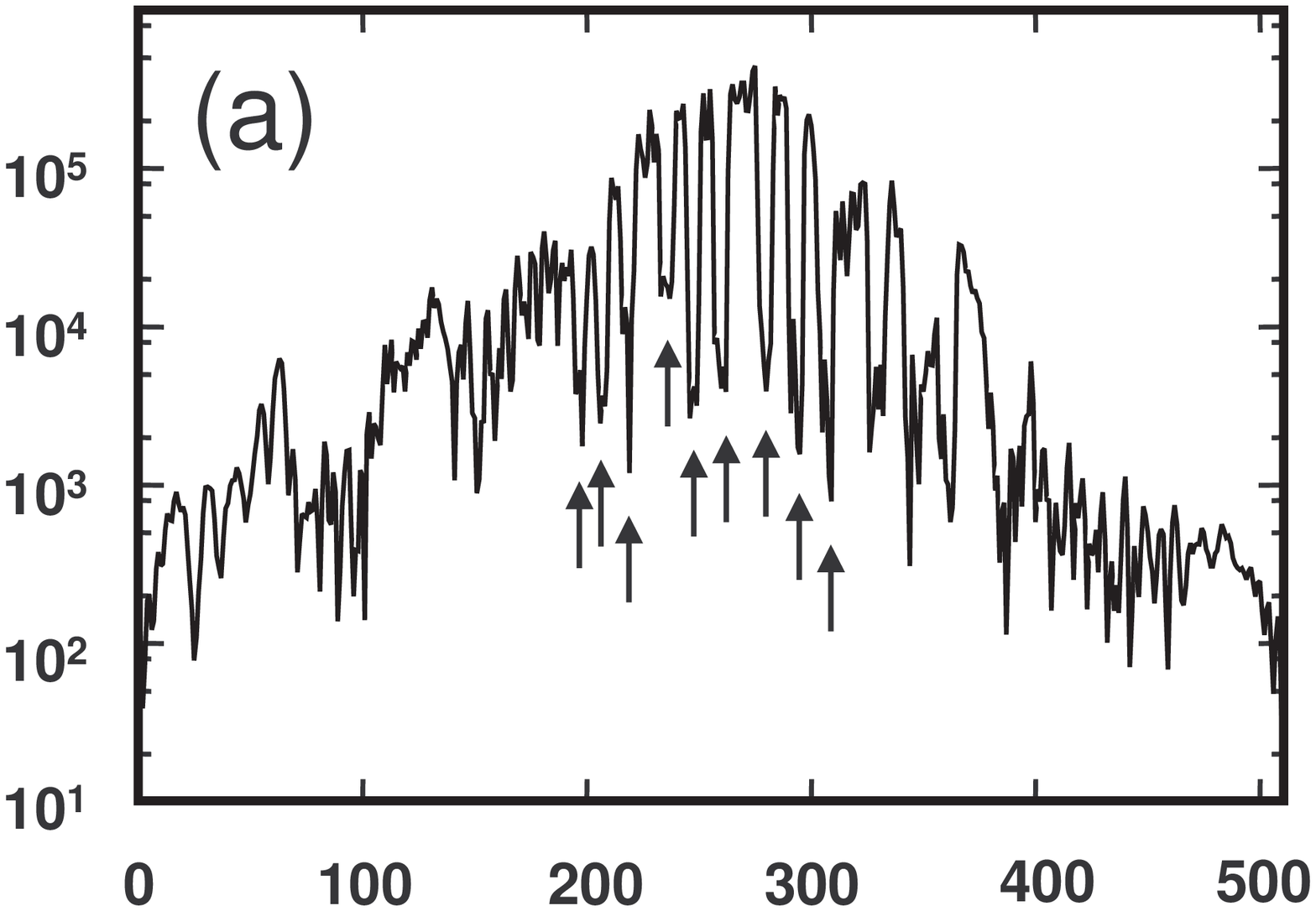}
\includegraphics[width = 4.0 cm,keepaspectratio=true]{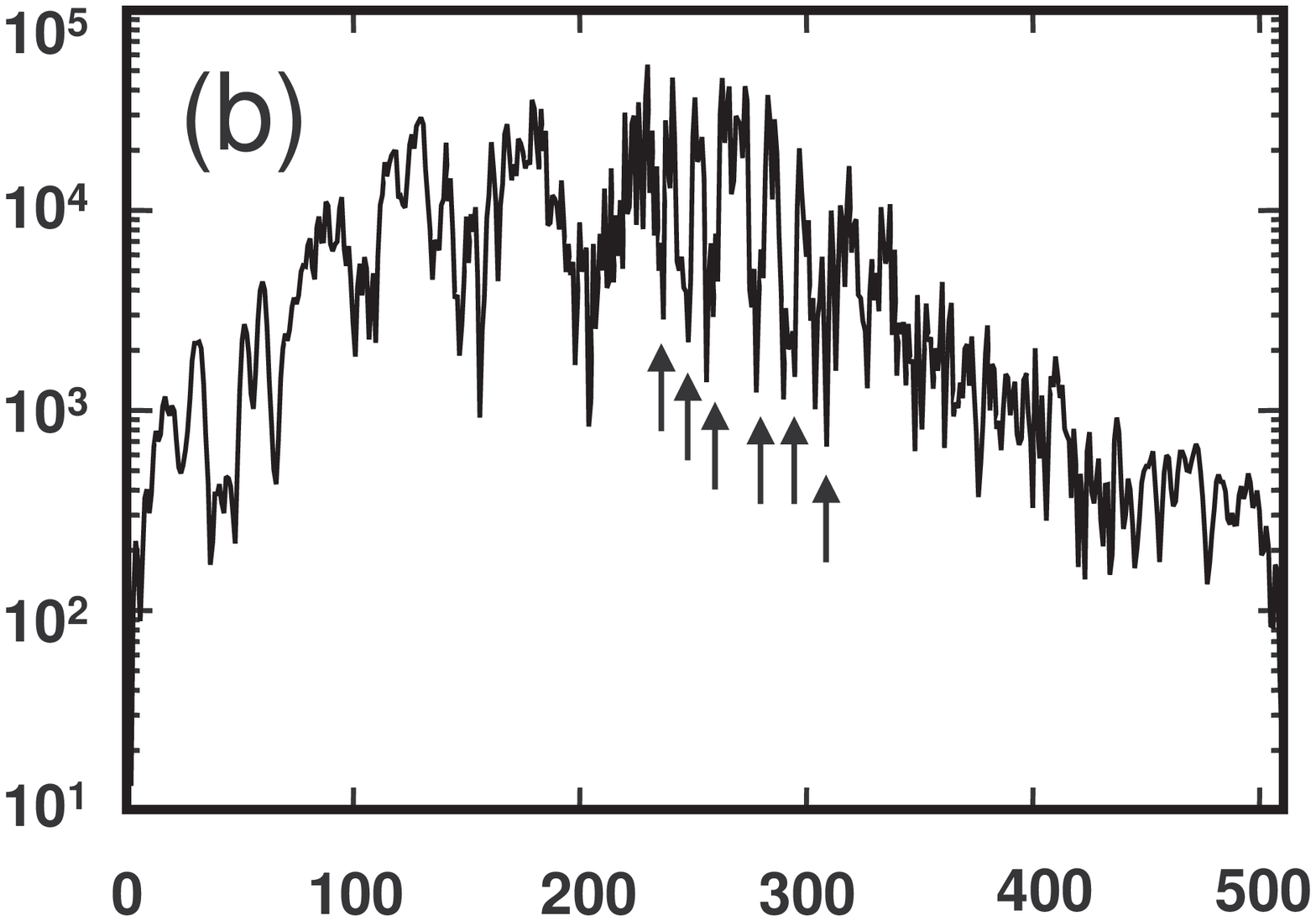}
\includegraphics[width = 4.0 cm,keepaspectratio=true]{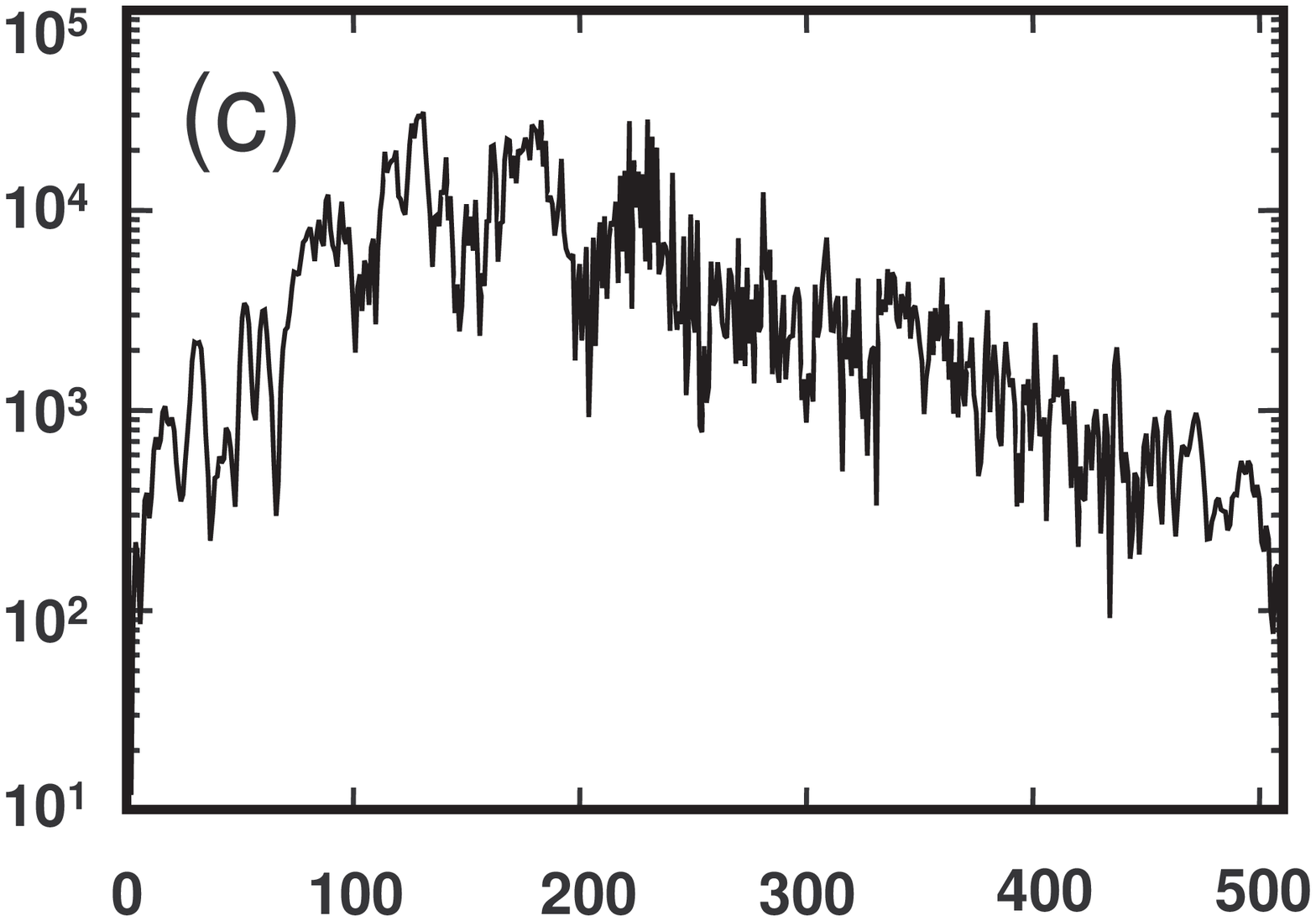}
\includegraphics[width = 4.0 cm,keepaspectratio=true]{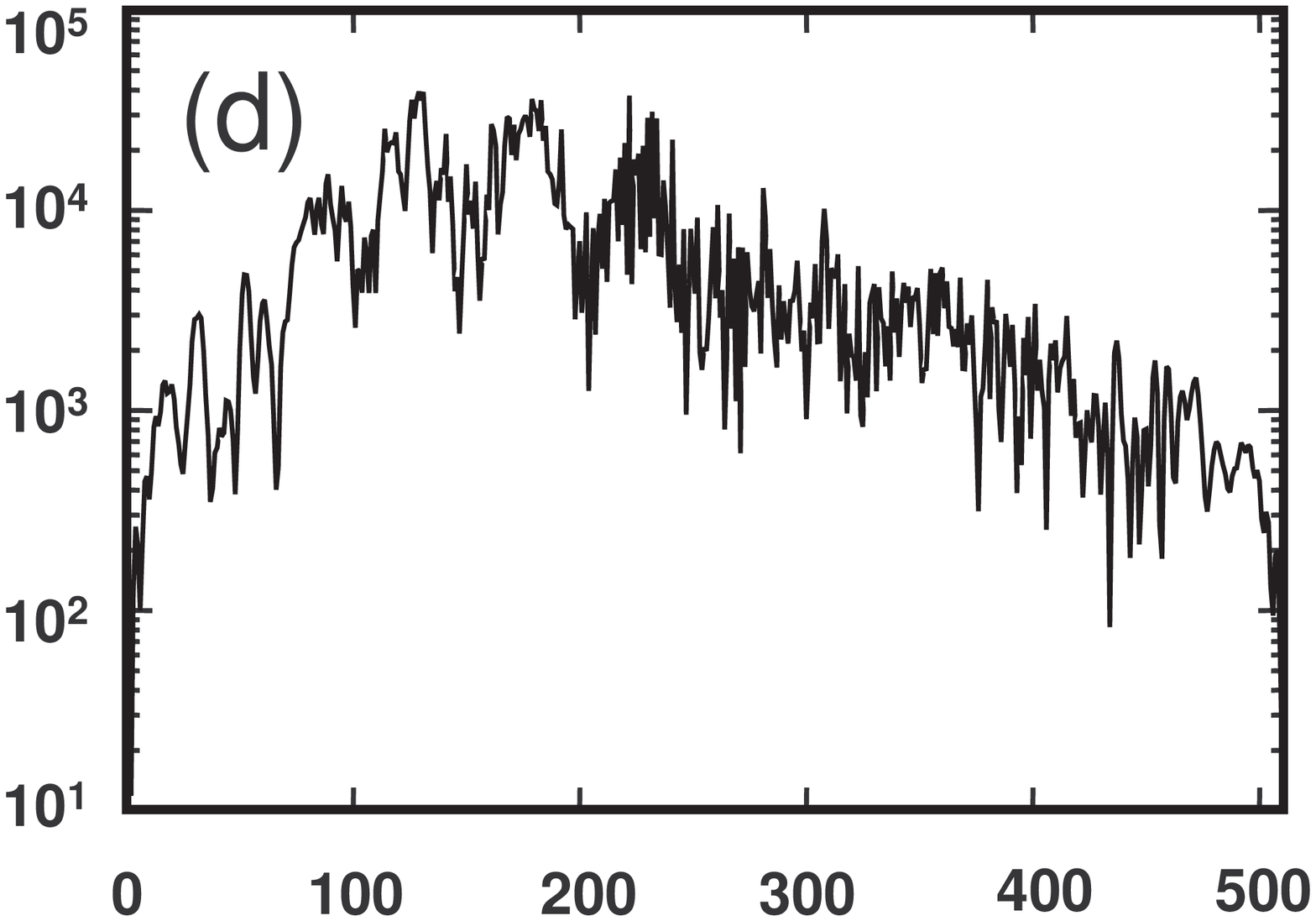}
\caption{ Horizontal cuts ($y \simeq 264 $) of the reconstructed
field intensity $|\tilde H|^2$ corresponding to the  images
Fig.\ref{fig_usaf_1ph}. (a) to (d) curves correspond to (a) to (d)
images. Vertical axis is $|\tilde H|^2$ in Arbitrary Units (A.U.).
Horizontal axis is pixel index $x=0...512$. Vertical display is
logarithmic. The total signal is $\simeq 4.3\times 10^8$ (a),
$8.7\times 10^6$ (b), $8.7 \times 10^4$ (c) and $1.2\times 10^4$
(d) photo electrons respectively for the sequence of 3 images.}
\label{fig_cut_usaf_1ph}
\end{center}
\end{figure}

We have performed a more quantitative study of the off-axis images
by performing cuts along the white dashed lines of
Fig.\ref{fig_usaf_1ph}. The curves, which are obtained by by
averaging over 3 pixels ($y=263$ to $y=265$), are displayed on
Fig.\ref{fig_cut_usaf_1ph}. On curves (a) and (b) the signal is
quite large ($\simeq 4.3\times 10^8$ and $8.7\times 10^6$ photo
electrons in (a) and (b)) and the USAF target black bars are visible
(see arrows on Fig.\ref{fig_cut_usaf_1ph}). Nevertheless, because of
the parasites, the 3 left hand side bars are not visible on curve
(b). On curves (c) and (d) with $8.7 \times 10^4$ and $1.2\times
10^4$ photo electrons respectively, the parasitic component is
dominant, and the USAF bars are not visible.

The curves give the quantitative weights of the signal and parasitic
components. Curves (b), (c) and (d)  have roughly the same shape
(except for the central region where the USAF bars are visible).
This means that, when the signal is less than $8.7\times 10^6$ photo
electrons (curve (b) ), most of the energy lies within parasitic
contributions.

\section{Phase-shifting holographic images}

As mentioned above, holographic images made from phase-shifting
measurements are reconstructed by using the whole set of 12
recorded CCD frames. Since the LO beam is phase shifted by $\pi/2$
between consecutive images (see Eq.\ref{Eq_nu_LO}), the object
complex hologram $H'$ is obtained by summing the CCD images with
the appropriate phase shift $\Delta \varphi = -m \pi/2$, where
$m=0...11$ is the image index:
\begin{equation}\label{equ_holo_4phi}
    H'(x,y)= \sum_{m = 0} ^{11}  (-j)^{m} ~I_m(x,y)
\end{equation}
\begin{eqnarray}\label{equ_holo_4phi_bis}
\nonumber H'= \sum_{m = 0} ^{11} (-j)^{m} \left(  \left| {\cal E}\right|^2 + \left| {\cal E}_{LO}\right|^2 \right)\\
\nonumber  ~~~~ + \sum_{m = 0} ^{11}  (-j)^{m}~ e^{+\omega_{CCD}t_m/4}~ {\cal E} {\cal E}_{LO}^*\\
 ~~~~+ \sum_{m = 0} ^{11} (-j)^{m}~ e^{-\omega_{CCD}t_m/4}~ {\cal E}^* {\cal E}_{LO}
\end{eqnarray}
By this choice of demodulation equation (Eq.\ref{equ_holo_4phi}),
the  zero order image  $\left| {\cal E}_{LO}\right|^2 $ and the twin
image ${\cal E}^* {\cal E}_{LO}$ terms are both zero, since $\sum_{m
= 0} ^{11} (-j)^m=0$, and  $ \sum_{m = 0} ^{11} (-j)^{m}
e^{-\omega_{CCD}t_m/4} \simeq 0$. Moreover, the true image ${\cal E}
{\cal E}_{LO}^*$ term is maximized, since $(-j)^{m}
e^{+\omega_{CCD}t_m/4}\simeq 1$. If the 4-phase condition is not
respected $\omega_L-\omega_{LO} \ne \omega_{CCD}/4 $, the twin image
term may  differ from zero. This does not greatly affect the final
result, since the zero order term, which is potentially much larger,
still cancels.

Like in single phase off-axis holography, phase-shifting holograms
are reconstructed by the convolution method that involves two FTs.
The $1280 \times 1024$ pixels wide $H'$ matrix is zero padded
within a square $2048 \times 2048$ calculation grid and the FTs
are calculated on this grid.

\begin{figure}[]
\begin{center}
\includegraphics[width = 8.2 cm,keepaspectratio=true]{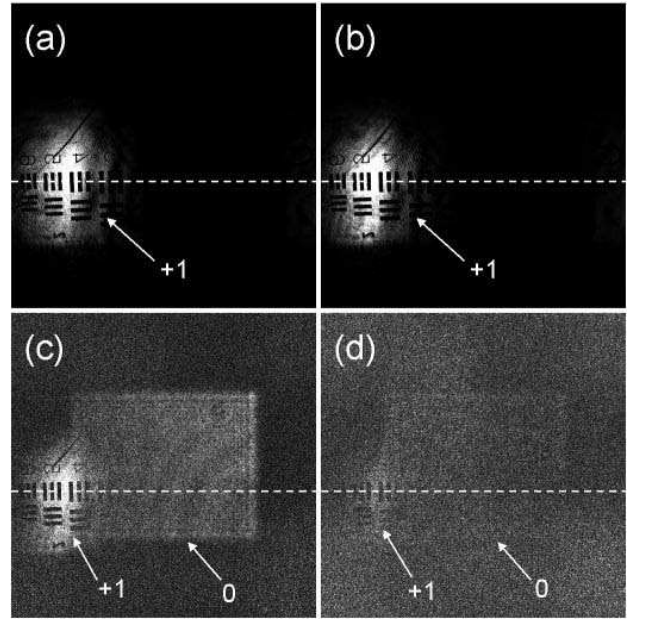}
\caption{4-phase reconstructed image of a USAF target in
transmission with low light illumination, without k-space
filtering. Images are reconstructed with $1.7 \times 10^9$ (a),
$3.5 \times 10^7$ (b), $3.5 \times 10^5$ (c) and $5 \times 10^4$
(d) photo electrons for all the pixels of the whole sequence of 12
images. $|H'|^2$ intensity displayed in linear gray scale.}
\label{fig_usaf_4ph}
\end{center}
\end{figure}

>From the same data collected to assess the sensitivity limit of the
single phase off-axis configuration, we have computed phase-shifting
holograms of the USAF target for different levels of illumination,
and we have reconstructed the images $H'(x,y,z=D)$ of the target
without any spatial filtering.

Fig.\ref{fig_usaf_4ph} shows the reconstructed images whose
intensity ($|H'(x,y,D)|^2$) is displayed in linear scale for $\simeq
1.7 \times 10^9$ (a), $3.5 \times 10^7$ (b), $3.5 \times 10^5$ (c)
and $5 \times 10^4$ (d) photo electrons for all the pixels and all
the 12 images of the sequence ($I_0...I_{11}$). These figures
correspond to the same attenuation level that are used in the single
phase off-axis case, but the signal, in photo electron units, is
four times larger, since the phase-shifting holographic images are
obtained by using 12 CCD images instead of 3 previously. Note that
the number of pixels used in the reconstruction calculation is
slightly larger than in reference \cite{gross_07} ($1280\times1024$
pixels instead of $1024\times1024$). As a consequence, for the same
experimental data, the total number of signal photo electrons must
be multiplied by a factor $\times 1.25$.

On Fig.\ref{fig_usaf_4ph}a and b, one see the USAF target with a
good $\textrm{SNR}$. Since the object beam is angularly tilted by
$\theta$ with respect to the LO beam, the USAF image is visible in
the left hand side of the reconstructed image domain, but,
contrarily to the Fig.\ref{fig_ima_1ph_700}c off-axis image, the
true image of the USAF target is visible without parasitic
contribution. Thanks to heterodyne phase-shifting
\cite{Leclerc2000}, the zero order and twin images are very low, and
are thus not visible. On Fig.\ref{fig_usaf_4ph}c, with $3.5 \times
10^5$ photo electrons, the USAF target is visible, but the zero
order image becomes visible too. On Fig.\ref{fig_usaf_4ph}d, with $5
\times 10^4$ photo electrons, the SNR is very low. One can only
guess the USAF target image in the left hand side of the image.

\begin{figure}[]
\begin{center}
\includegraphics[width = 4. cm,keepaspectratio=true]{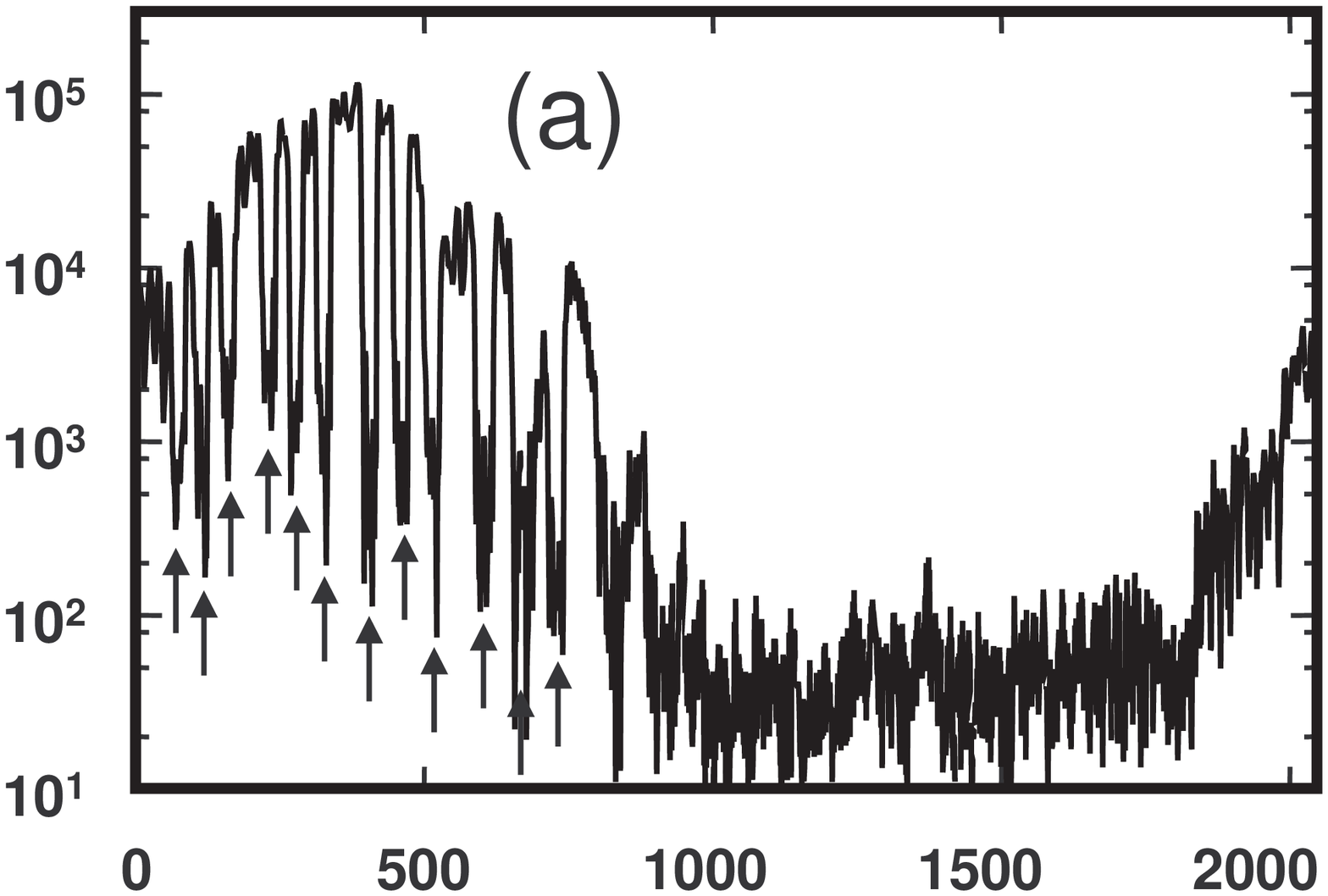}
\includegraphics[width = 4. cm,keepaspectratio=true]{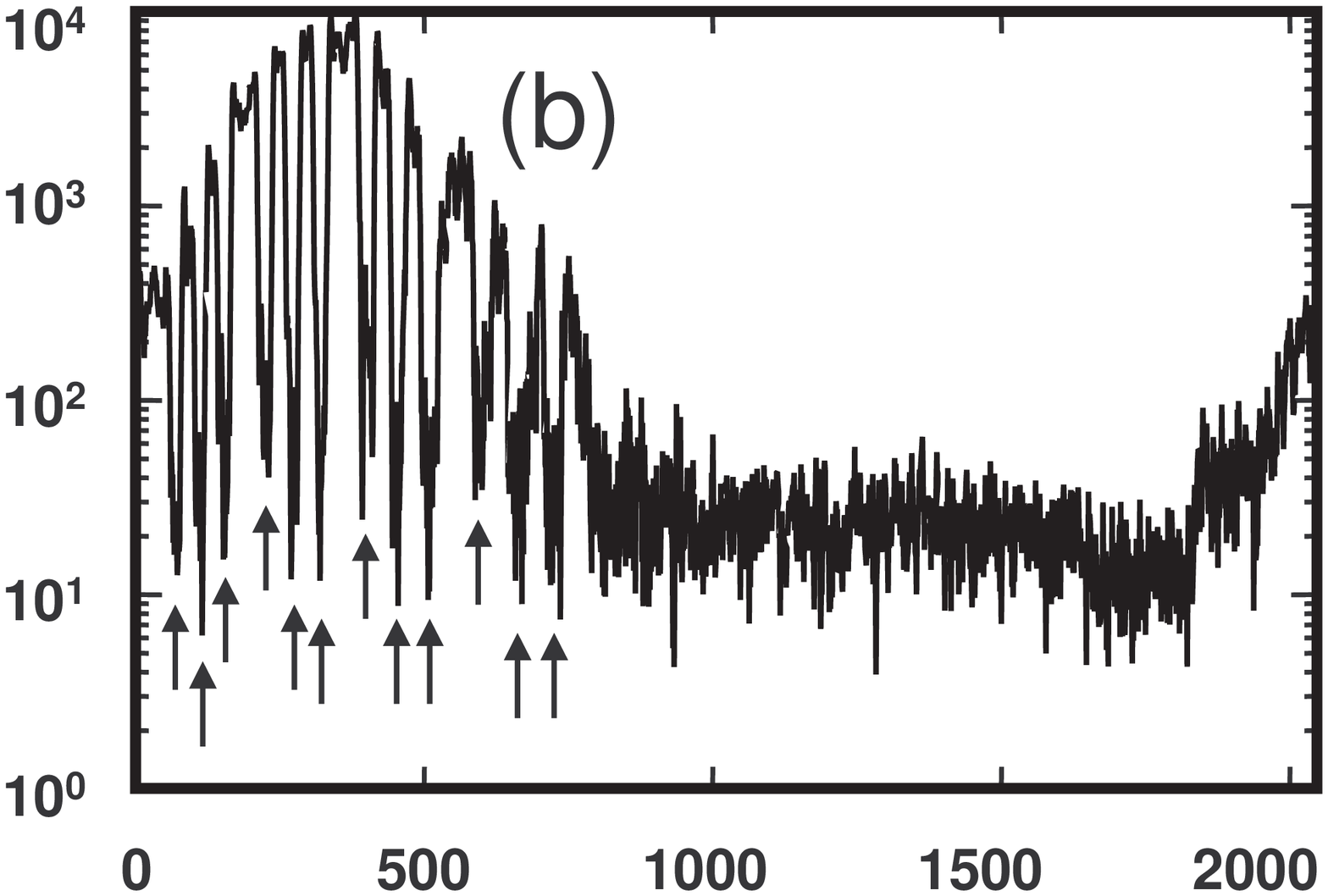}
\includegraphics[width = 4. cm,keepaspectratio=true]{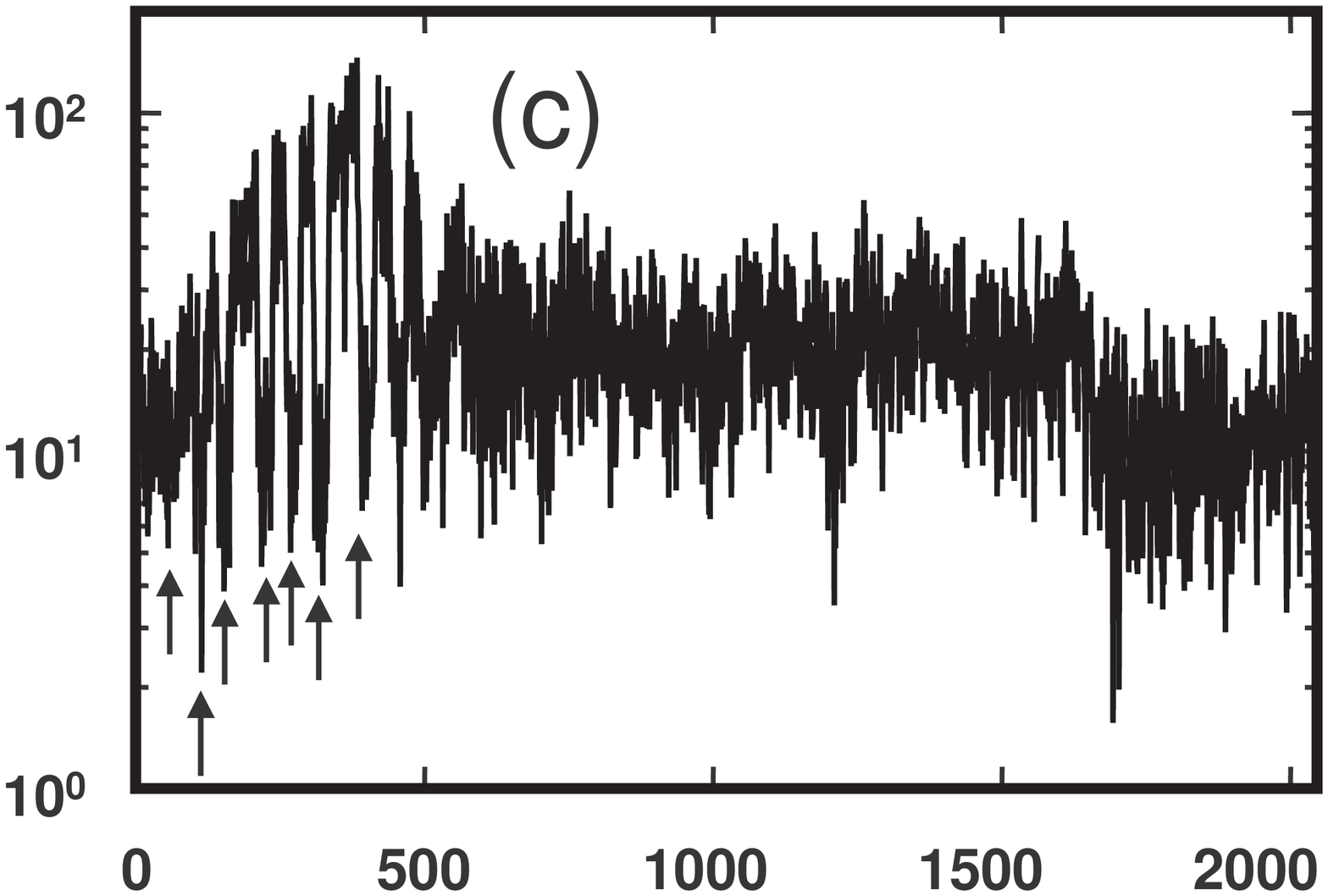}
\includegraphics[width = 4. cm,keepaspectratio=true]{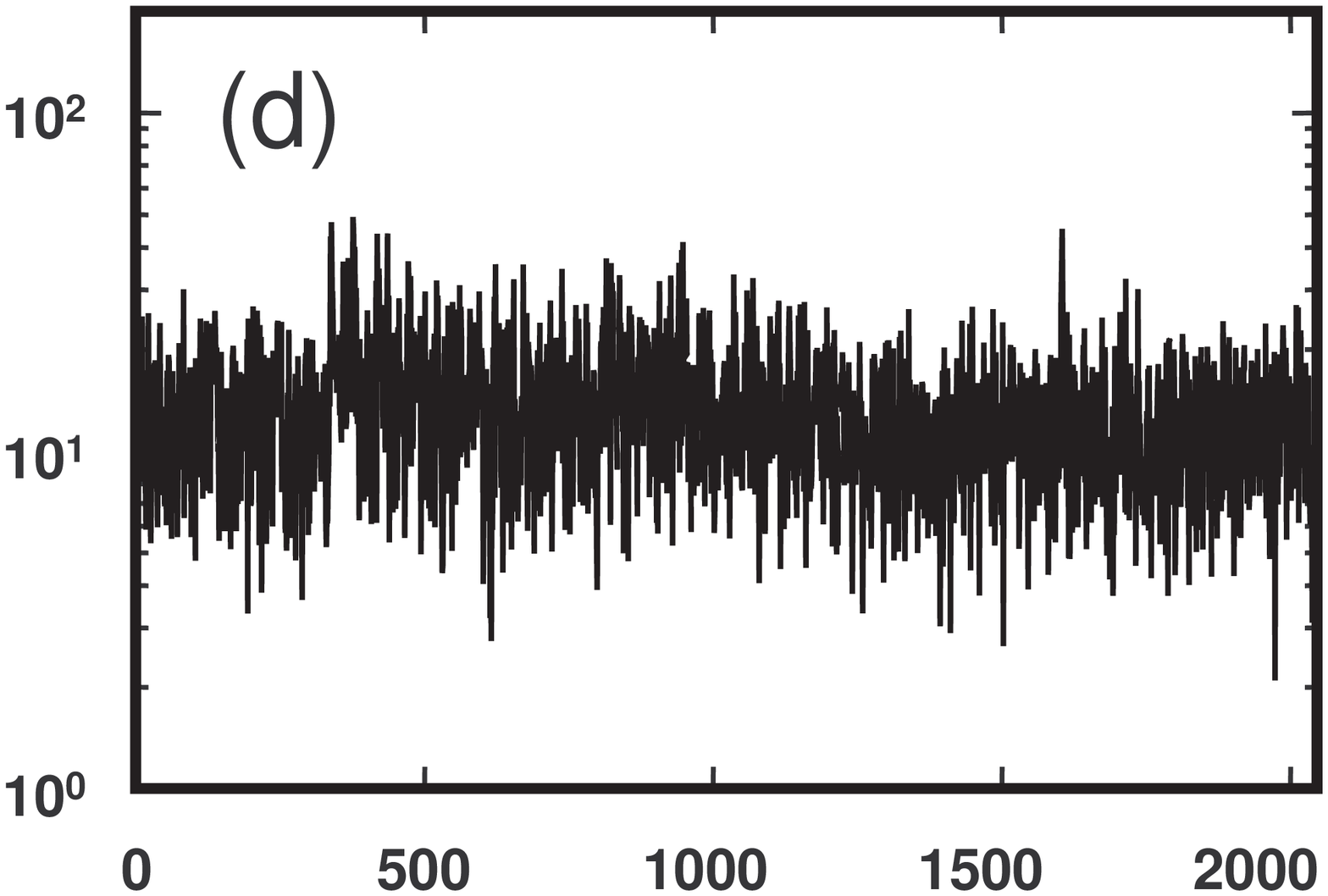}
\caption{ Horizontal cuts ($y \simeq 1196 $) of the reconstructed
field intensity $|\tilde H|^2$ corresponding to the 4 phase images
of Fig.\ref{fig_usaf_4ph}. (a) to (d) curves correspond to (a) to
(d) images. Vertical axis is $|\tilde H|^2$ in Arbitrary Units
(A.U.). Horizontal axis is pixel index $x=0...2048$. Vertical
display is logarithmic. The total signal is $\simeq 1.7 \times 10^9$
(a), $3.5 \times 10^7$ (b), $3.5 \times 10^5$ (c) and $5 \times
10^4$ (d)   photo electrons for all the pixels of the whole sequence
of 12 images.} \label{fig_cut_usaf_4ph}
\end{center}
\end{figure}

The curves that are obtained  by performing an horizontal cut ($y
\simeq 1196$) and by averaging the signal intensity $|H|^2$ over
11 pixels ($y=1191$ to $y=1201$), are displayed on
Fig.\ref{fig_cut_usaf_4ph}. On curves (a), (b) and (c) with
$\simeq 1.7 \times 10^9$, $3.5 \times 10^7$ and $3.5 \times 10^5$
photo electrons the USAF target black bars are clearly visible in
the left hand side of the curves (see arrows on
Fig.\ref{fig_cut_usaf_4ph}). On curves (d) the signal is lower
than the zero order image background and the  bars are not
visible.

Heterodyne phase-shifting is thus very efficient for getting good
images at low illumination level. Phase-shifting holography is
clearly better in our case than off-axis holography with or without
spatial filtering. By making the difference of images in
Eq.\ref{equ_holo_4phi} the $|{\cal E}_{LO}|^2$ term is cancelled,
and most of the LO beam contribution to noise and parasitic signal
is removed by phase-shifting holography. This is especially
important at low illumination level, since, in that case, the LO
beam, which brings noise and parasites, is of much larger power than
the signal.

\begin{figure}[]
\begin{center}
\includegraphics[width = 8.2 cm,keepaspectratio=true]{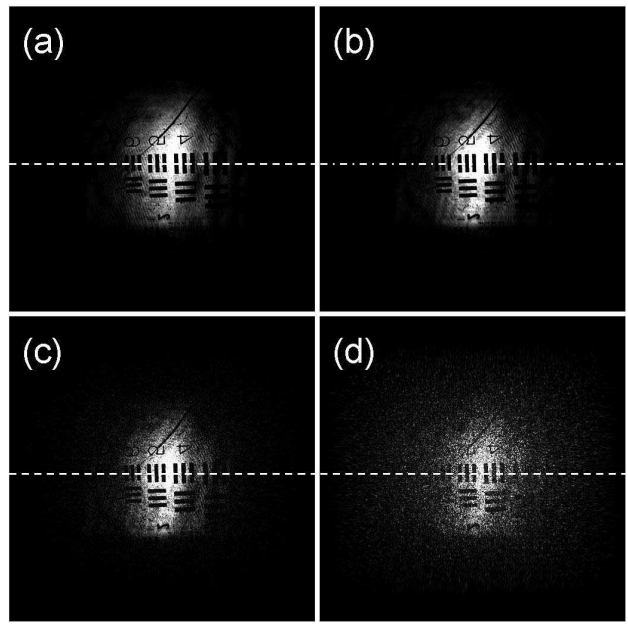}
\caption{4-phase reconstructed image of a USAF target in
transmission at low light illumination with k-space filtering.
Images are reconstructed with $\simeq 1.7 \times 10^9$ (a), $3.5
\times 10^7$ (b), $3.5 \times 10^5$ (c) and   $5 \times 10^4$ (d)
photo electrons for all the pixels of the whole sequence of 12
images. $|H'|^2$ intensity displayed in linear gray scale.}
\label{fig_usaf_4ph_bis}
\end{center}
\end{figure}

Since the phase-shifting holograms have been recorded in off-axis
geometry, one can use the spatial filtering technique to improve
further the quality of the reconstructed images, as done in
reference \cite{gross_07}. We have selected in the k-space a $400
\times 400$ region centered on the true image, copied this region in
the center of a $512\times 512$ calculation grid (zero padding) and
then calculated the USAF target images.

The reconstructed images are displayed on
Fig.\ref{fig_usaf_4ph_bis} for the same levels of illumination
that for Fig.\ref{fig_usaf_4ph}. Since the true image selected
zone is translated in the center of the k-space domain, the USAF
target is seen in the center of the reconstructed images. At high
level of illumination, for $\simeq 1.7 \times 10^9$ (a) and  $3.5
\times 10^7$ (b)  photo electrons, the USAF images are seen with
high SNR.  The spatial filtering method does not seem to improve
the image quality. On Fig.\ref{fig_usaf_4ph_bis} (c), with $3.5
\times 10^5$ (c) photo electrons,  the USAF target is still seen
with a good $\textrm{SNR}$. One can see that the image quality is
better with spatial filtering (Fig.\ref{fig_usaf_4ph_bis} (c))
than without (Fig.\ref{fig_usaf_4ph} (c)). Spatial filtering
lowers the number of modes that bring noise, and cancels the zero
order image that is seen on Fig.\ref{fig_usaf_4ph} (c) (arrow 0,
rectangular region). On Fig.\ref{fig_usaf_4ph_bis} (d), with $5
\times 10^4$ (c) photo electrons,  the USAF target is still seen
with $\textrm{SNR} \sim 1$. As explained in reference
\cite{gross_07}, the Fig.\ref{fig_usaf_4ph_bis} (d) image is
obtained with about one photo electron per resolved pixel of the
reconstructed image (or per k-space mode), for the whole sequence
of 12 images.

\begin{figure}[]
\begin{center}
\includegraphics[width = 4. cm,keepaspectratio=true]{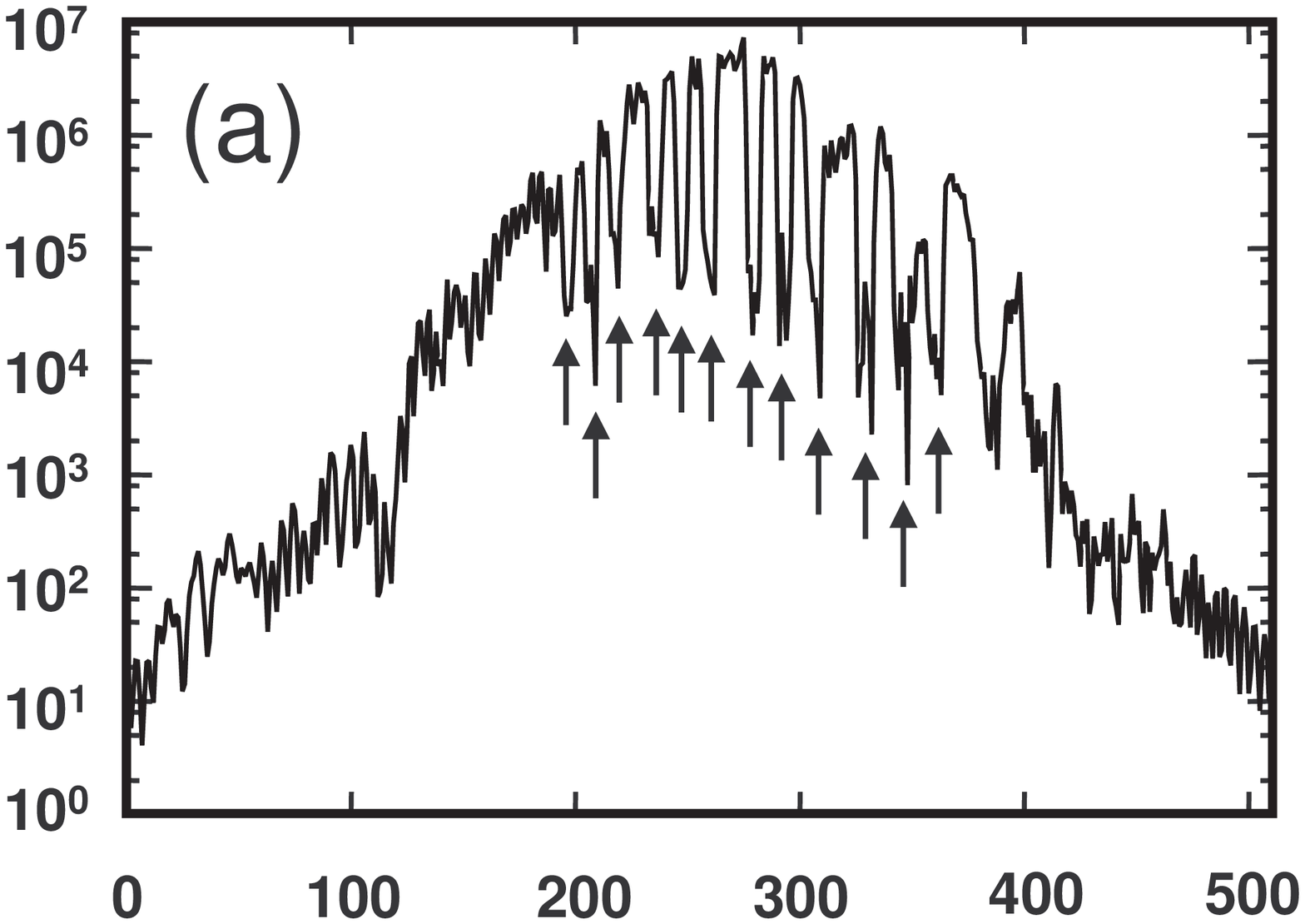}
\includegraphics[width = 4. cm,keepaspectratio=true]{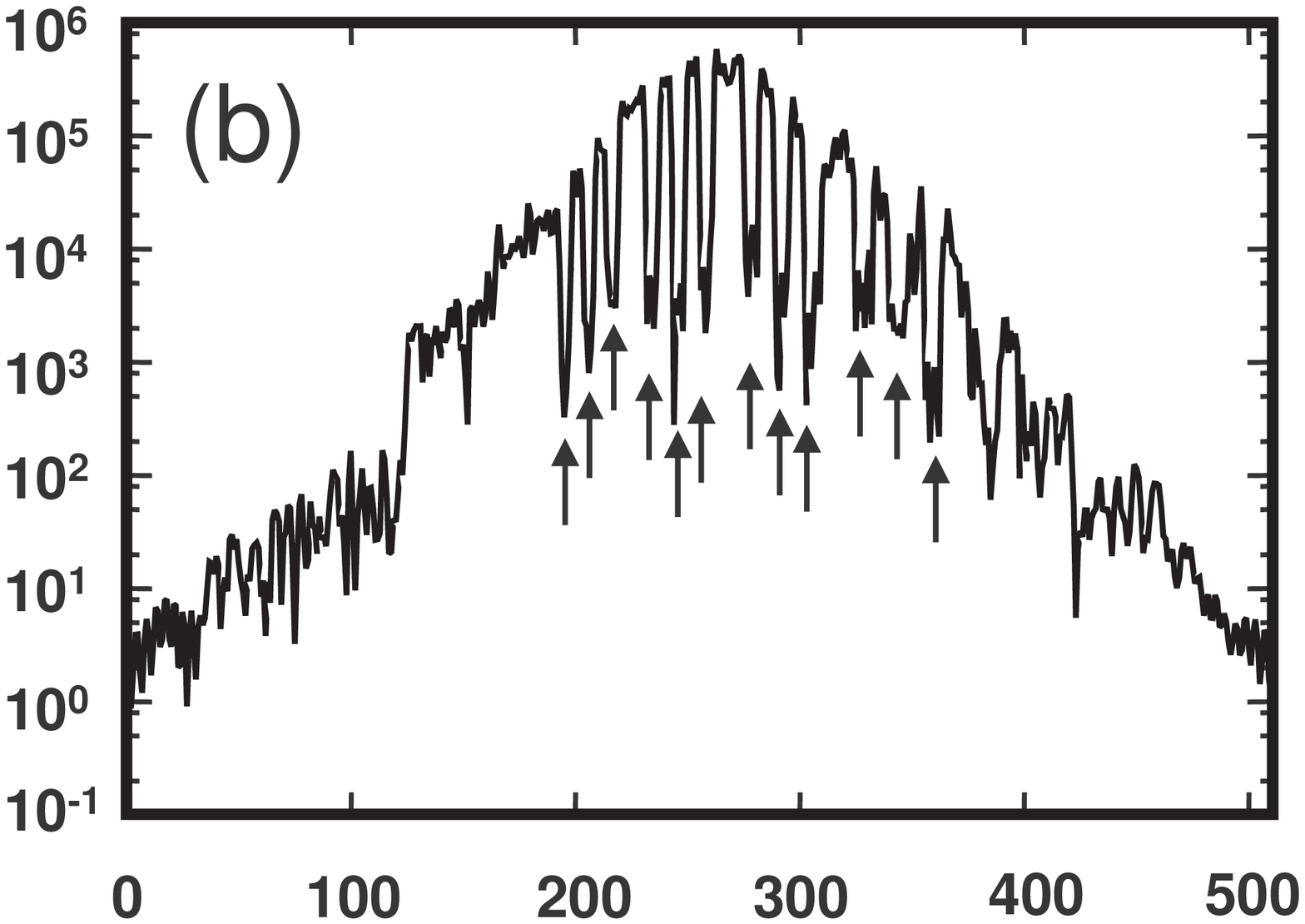}
\includegraphics[width = 4. cm,keepaspectratio=true]{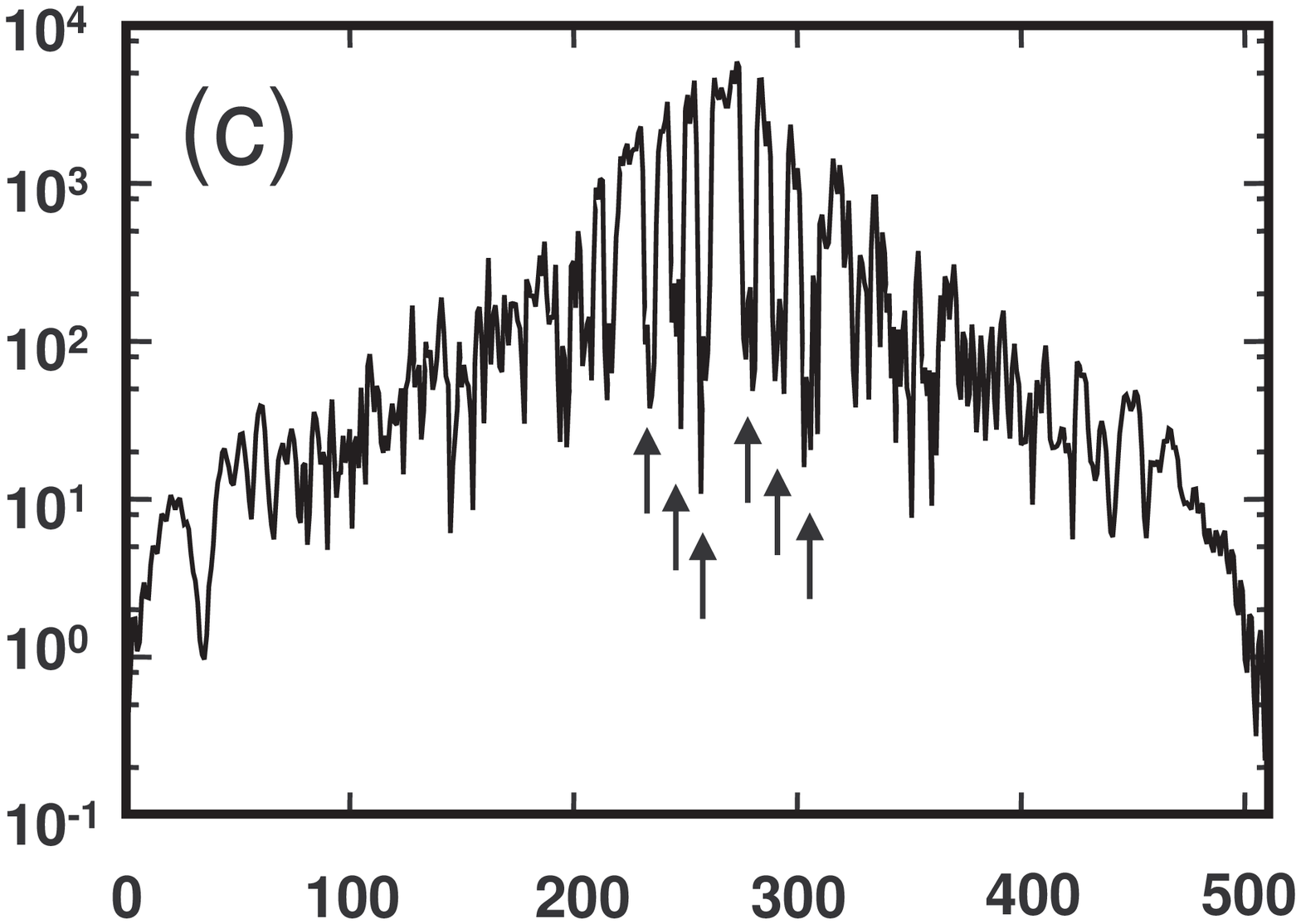}
\includegraphics[width = 4. cm,keepaspectratio=true]{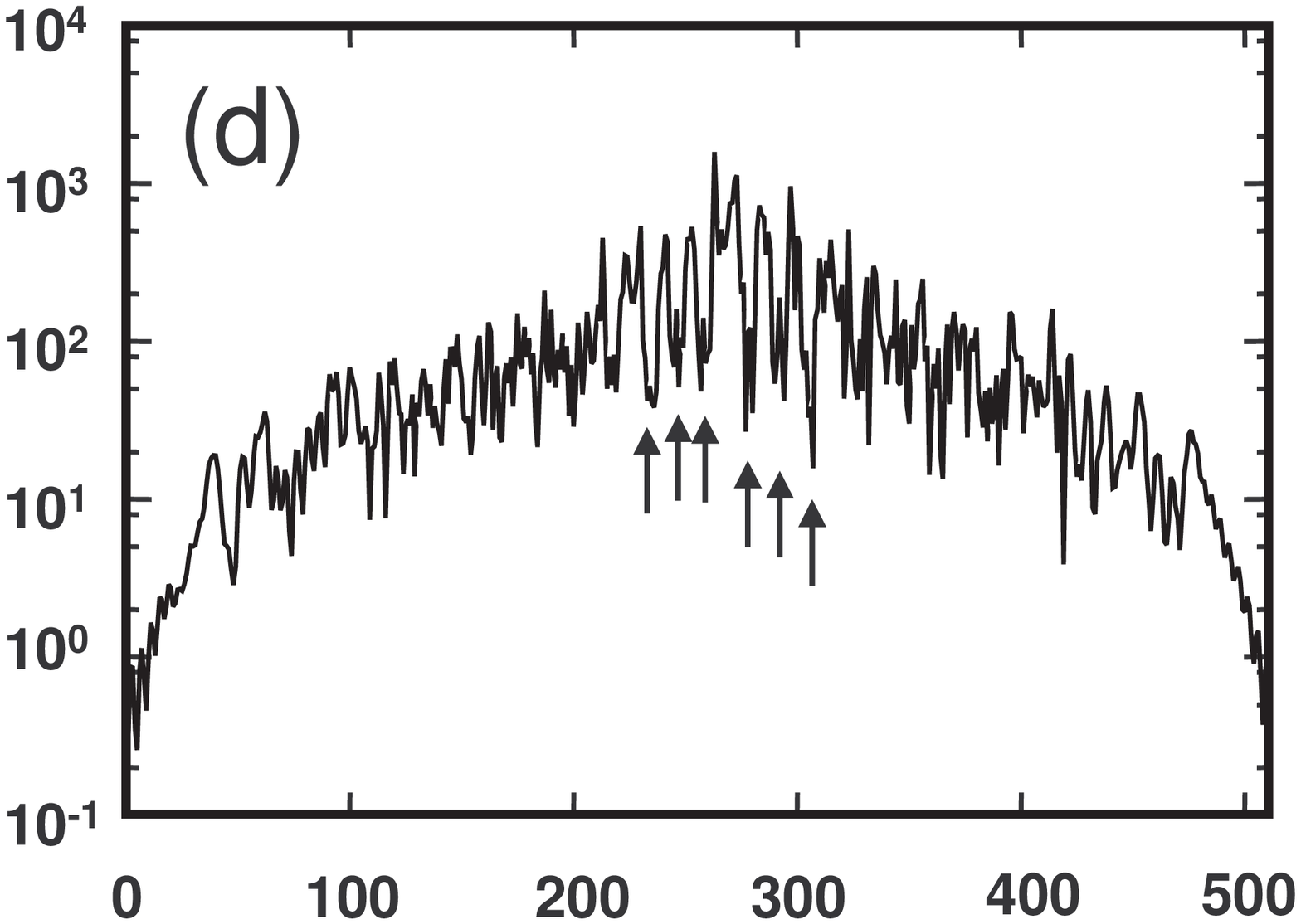}
\caption{Horizontal cuts ($y \simeq 264 $) of the reconstructed
field intensity $|\tilde H|^2$ corresponding to the  images
Fig.\ref{fig_usaf_4ph_bis}. (a) to (d) curves correspond to (a) to
(d) images. Vertical axis is $|\tilde H|^2$ in Arbitrary Units
(A.U.). Horizontal axis is pixel index $x=0...512$. Vertical
display is logarithmic. The total signal is $\simeq 1.7 \times
10^9$ (a), $3.5 \times 10^7$ (b), $3.5 \times 10^5$ (c) and   $5
\times 10^4$ (d) photo electrons for all the pixels of the whole
sequence of 12 images.} \label{fig_cut_usaf_4ph_bis}
\end{center}
\end{figure}

The curves, which are obtained by averaging over 3 pixels ($y=263$
to $y=265$), are displayed on Fig.\ref{fig_cut_usaf_4ph_bis}. As
seen, the USAF target black bars (see arrows) are easily visible
on all curves. Image and cuts of Fig.\ref{fig_usaf_4ph_bis} and
Fig.\ref{fig_cut_usaf_4ph_bis}, which have been obtained with the
combination of techniques of  reference \cite{gross_07} can be
considered here as the optimal results to be reach.

%

\section{Discussion}

Let us discuss on the respective merits of the techniques that are
combined in reference \cite{gross_07} in order to get a
shot-noise-limited holographic measurement.

As seen on the Fig.\ref{fig_ima_1ph_700} (c) reconstructed image,
and on the Fig.\ref{fig_cut_1ph_700} cut, which are obtained with a
large signal of $4.3 \times 10^8$ photo electrons, the LO beam zero
order alias is much brighter than the signal itself. This is
confirmed by Fig.\ref{fig_cut_1ph_700}. On the  cut, the zero order
alias (pixel 500 to 1600 plateau), is about 20 times larger that the
signal itself (peeks and valleys marked by arrows, in the pixel 0 to
300 region).  Moreover, the image and the zero order alias overlap.
In single-phase regime, it is thus necessary to use a spatial
filtering technique, as done on Fig.\ref{fig_cut_usaf_1ph} (a), in
order to see the object without zero order alias overlapping. Such
spatial filtering is nevertheless far to be sufficient to reach the
shot noise limit with our experimental data, as shown by
Fig.\ref{fig_usaf_1ph} (c,d) and Fig.\ref{fig_cut_usaf_1ph} (c,d).
By comparison, the 4-phase method without spatial filtering yields
much better results. This is illustrated by comparing the
Fig.\ref{fig_usaf_1ph} and Fig.\ref{fig_cut_usaf_1ph} images and
cuts, obtained with 1-phase holograms and spatial filtering,  with
the corresponding Fig.\ref{fig_usaf_4ph} and
Fig.\ref{fig_cut_usaf_4ph} images and cuts, obtained with 4-phase
holograms, but without spatial filtering.


In the case of the 1-phase images and cuts presented on
Fig.\ref{fig_usaf_1ph} and Fig.\ref{fig_cut_usaf_1ph}, the results
are far from the shot-noise limit because of the parasites, visible
on Fig.\ref{fig_ima_1ph_700} (d) (arrow 2). We have done many
experiments with the setup sketched in Fig.\ref{fig_setup_usaf}.
With 1-phase, parasitic contributions are almost ever there, but
their location depends on the  beam splitter cube (BS) orientation.
We guess that parasites are related to unwanted LO beam reflections
on the BS, whose faces are not perfectly parallel.

\begin{figure}[]
\begin{center}
\includegraphics[width = 8.2 cm,keepaspectratio=true]{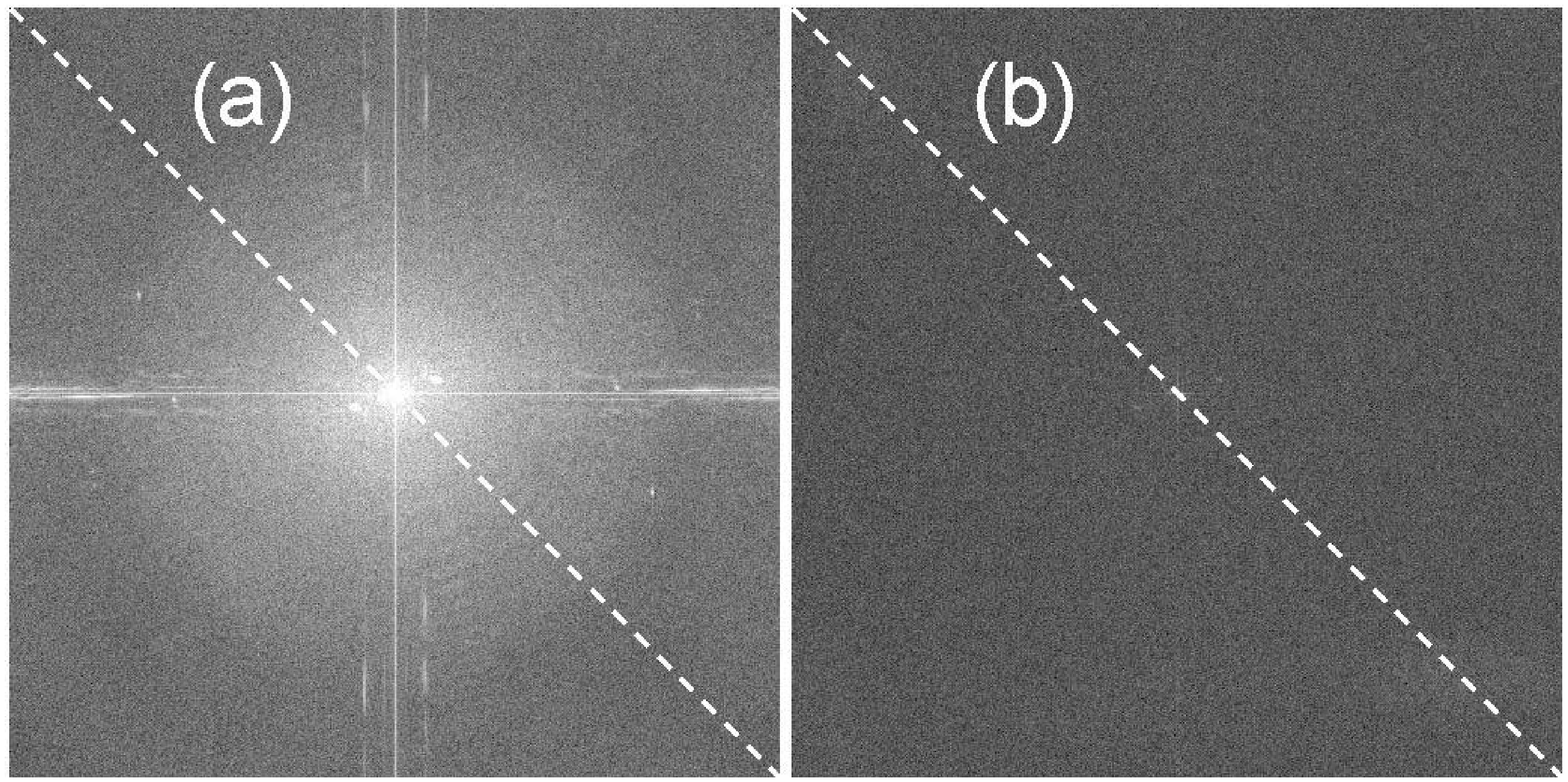}
\includegraphics[width = 4. cm,keepaspectratio=true]{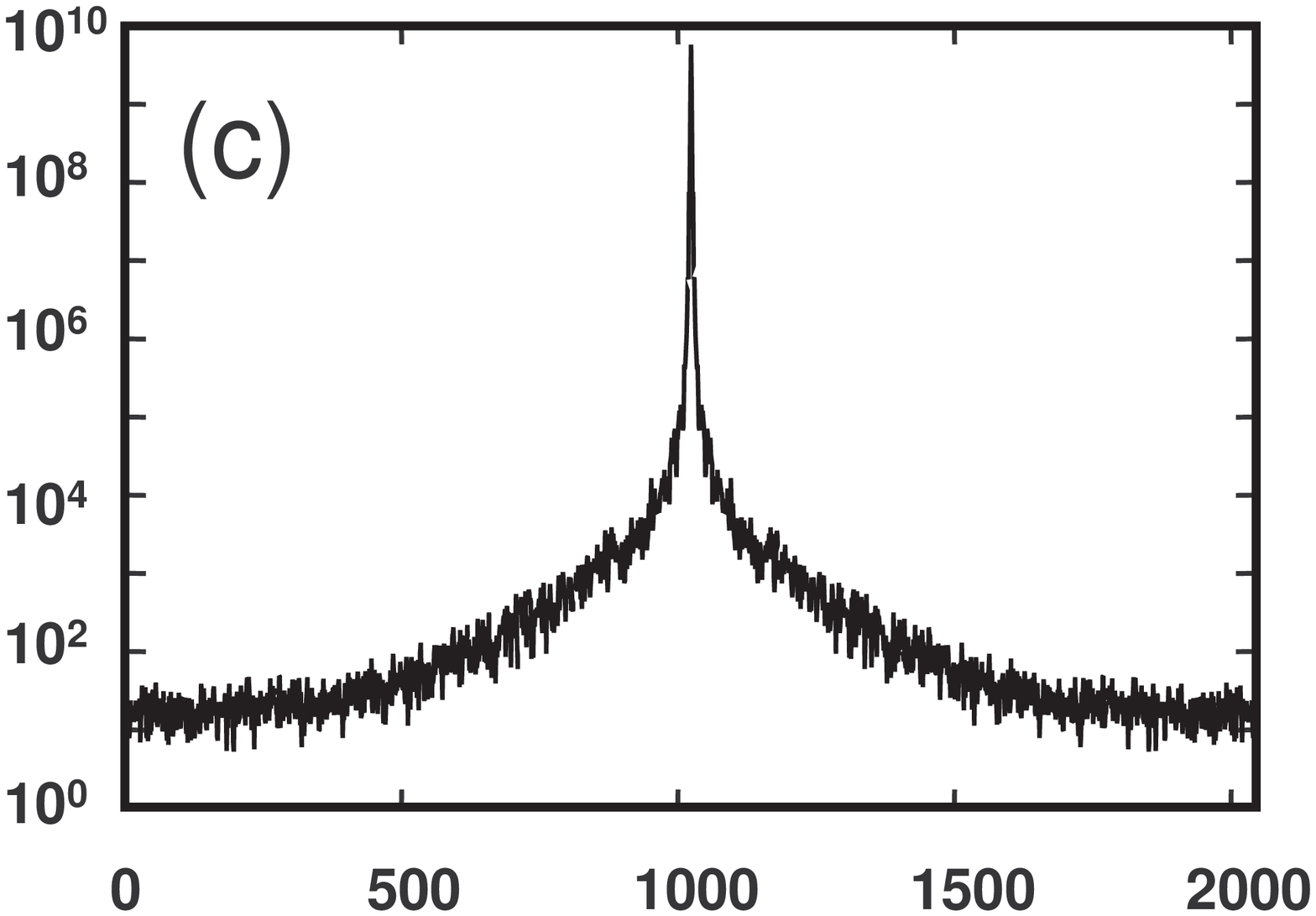}
\includegraphics[width = 4. cm,keepaspectratio=true]{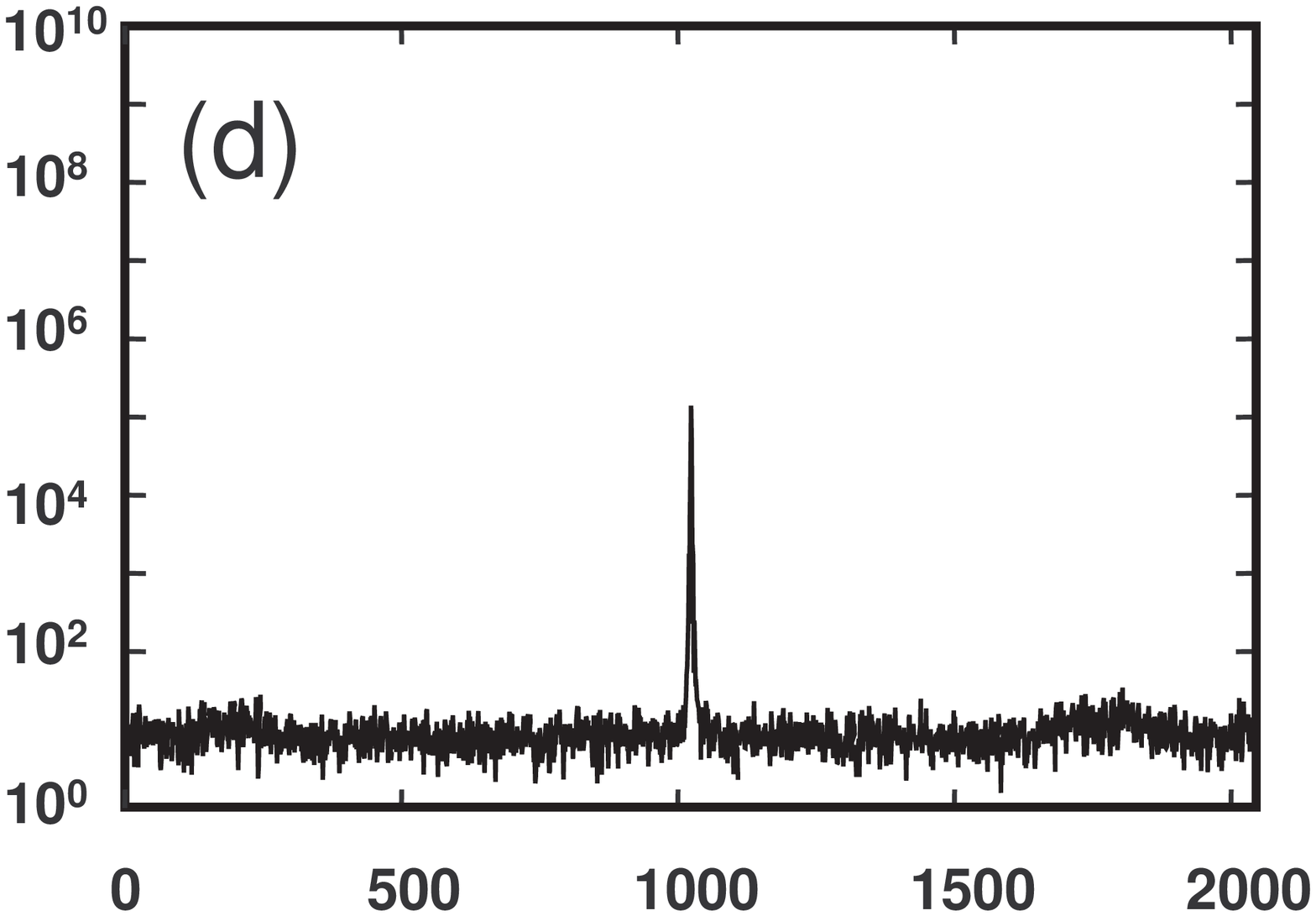}
\caption{(a,b) k-space field intensity $|\tilde H|^2$ and $|\tilde
H'|^2$ reconstructed from 1-phase (a) and a 4-phase  holograms
obtained without signal (i.e. without illumination of the USAF
char). (c,d) cut along the white dashed diagonal lines for 1-phase
(c) and 4-phase (d) holograms. Vertical axis is k-space intensity
$|\tilde H|^2$ and $|\tilde H'|^2$ along cut. Vertical axis is pixel
index. Intensity is averaged over 11 pixels i.e. in the interval
$(i,i-5)...(i,+5)$ where $i=0...2047$ is the pixel index.}
\label{fig_kspace_image_cut_zero}
\end{center}
\end{figure}
To improve the quality of the 1-phase images, one can modify the BS
orientation, such the k-space image of the object is moved in a
"quiet zone" of the k-space, i.e. a zone without parasitic alias,
like the upper right corner of the Fig.\ref{fig_ima_1ph_700} (b)
k-space image.  But this is not sufficient to reach the shot noise
limit. To illustrate this point, we have extracted, in the reference
\cite{gross_07} experimental data, a sequence of 12 images without
signal beam (i.e. with the LO beam alone). With these data, we have
computed 1-phase and 4-phase holograms (Eq.\ref{equ_holo} and
Eq.\ref{equ_holo_4phi}, respectively). The k-space intensity images,
$|\tilde H(k_x,k_y)|^2 $ and $|\tilde H'(k_x,k_y)|^2 $, are
displayed  on Fig.\ref{fig_kspace_image_cut_zero} (a) and (b), with
the same logarithmic gray scale.

With 1-phase (a), the LO beam image is much brighter, and extend
over a much larger area, than in the 4-phase case (b). To make a
quantitative analysis of these images, we have plotted profiles
along the Fig.\ref{fig_kspace_image_cut_zero} (a) and (b) white
diagonal dashed lines. Note that we have made diagonal cuts in order
to explore  "quiet zones" of the k-space. The intensity signal
($|\tilde H|^2$ or $|\tilde H'|^2$) along the cut is represented on
Fig.\ref{fig_kspace_image_cut_zero} (c) and (d). For $k_x=k_y=0$, we
get a peak on the (c) and (d) cuts, which corresponds to the flat
field  component of the LO field.  In the 1-phase case, the peak is
about $\simeq 4\times 10^4$ larger than in the 4-phase case. The
1-phase peak is also much broader, so that in most of the k-space
domain, the 1-phase LO parasitic signal (c) is several orders of
magnitude larger than its 4-phase counterpart (d). It is thus much
larger than the shot noise limit, which is equal to the 4-phase
noise floor (within a few per cent, as  verified experimentally).
The shot noise limit is thus not reached with 1-phase detection.
This means that the LO beam signal cannot be fully filtered-off by
spatial filtering. We guess that, in real life experiments, it is
extremely difficult to have a perfect flat field LO beam. Thus,
complete spatial filtering the LO beam cannot be achieved.

\begin{figure}[]
\begin{center}
\includegraphics[width = 8.2 cm,keepaspectratio=true]{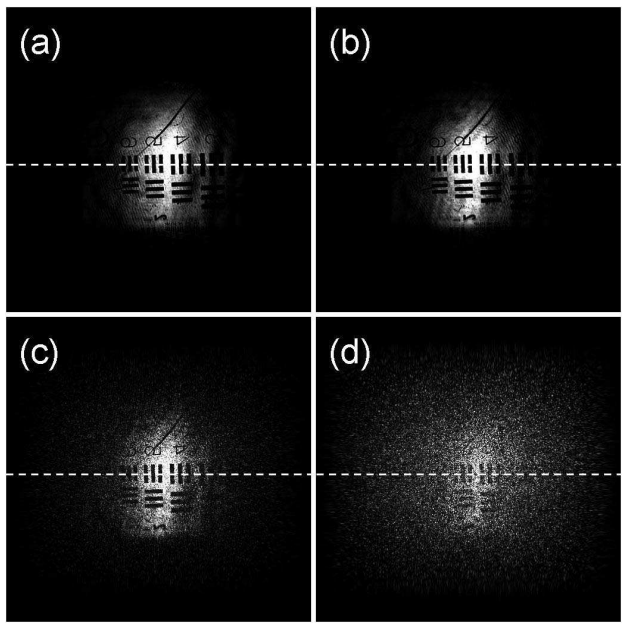}
\caption{1-phase reconstructed images of a USAF target in
transmission with substraction of  the average LO beam signal.
Images are obtained with k-space filtering with $\simeq 4.3\times
10^8$ (a), $8.7\times 10^6$ (b), $8.7 \times 10^4$ (c) and
$1.2\times 10^4$ (d) photo electron respectively for the sequence
of 3 images.} \label{fig_usaf_14ph_f}
\end{center}
\end{figure}
\begin{figure}[]
\begin{center}
\includegraphics[width = 4. cm,keepaspectratio=true]{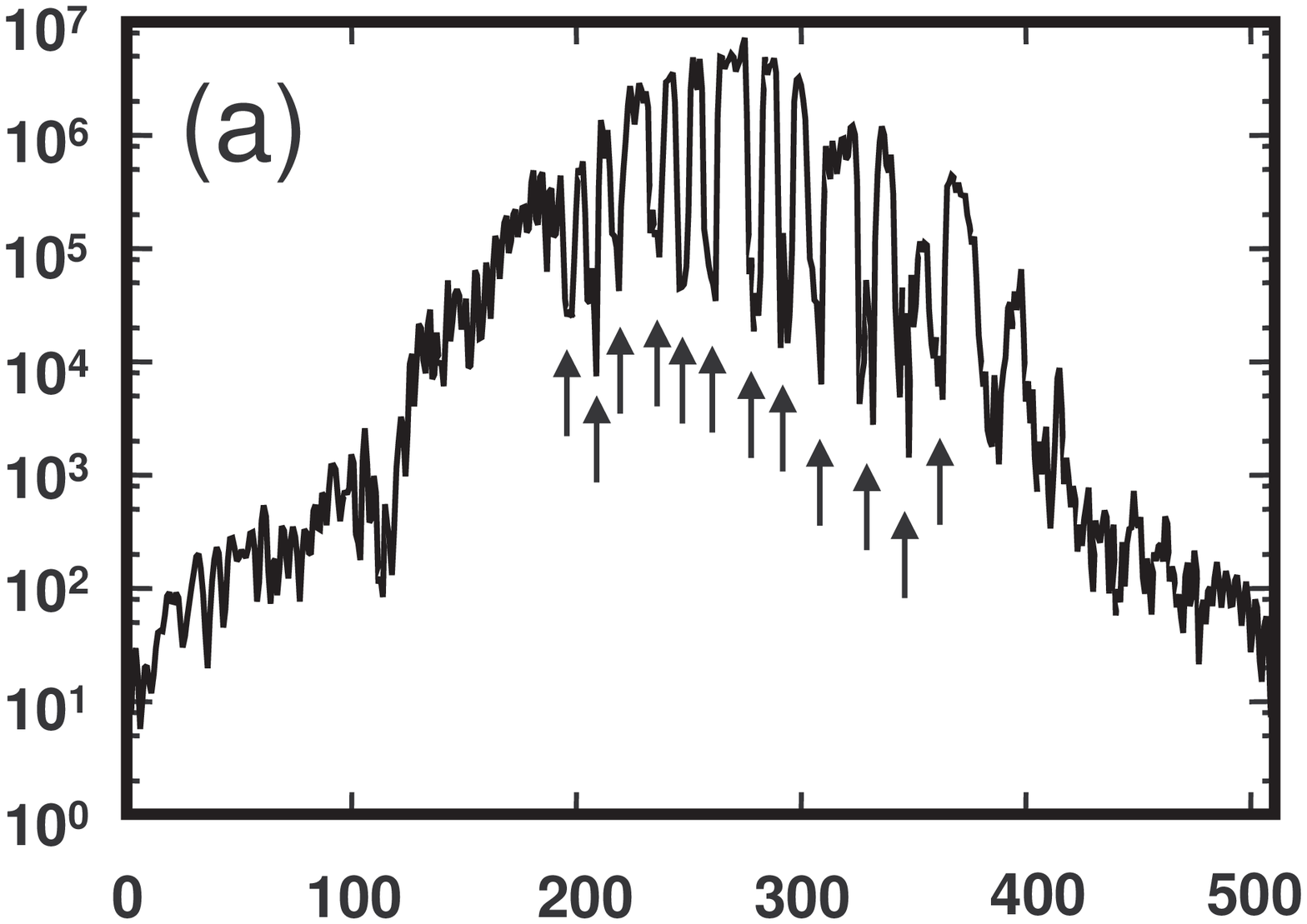}
\includegraphics[width = 4. cm,keepaspectratio=true]{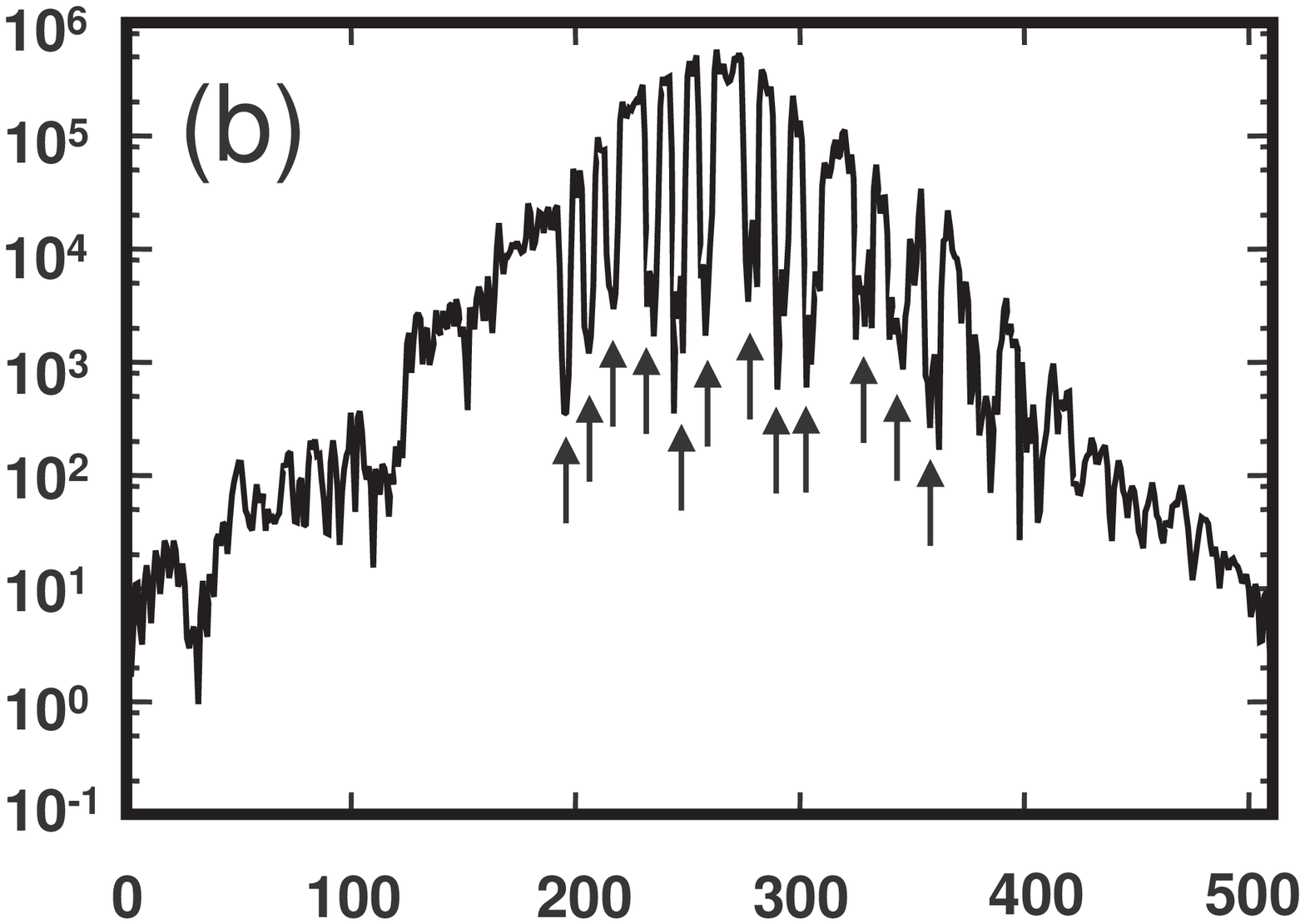}
\includegraphics[width = 4. cm,keepaspectratio=true]{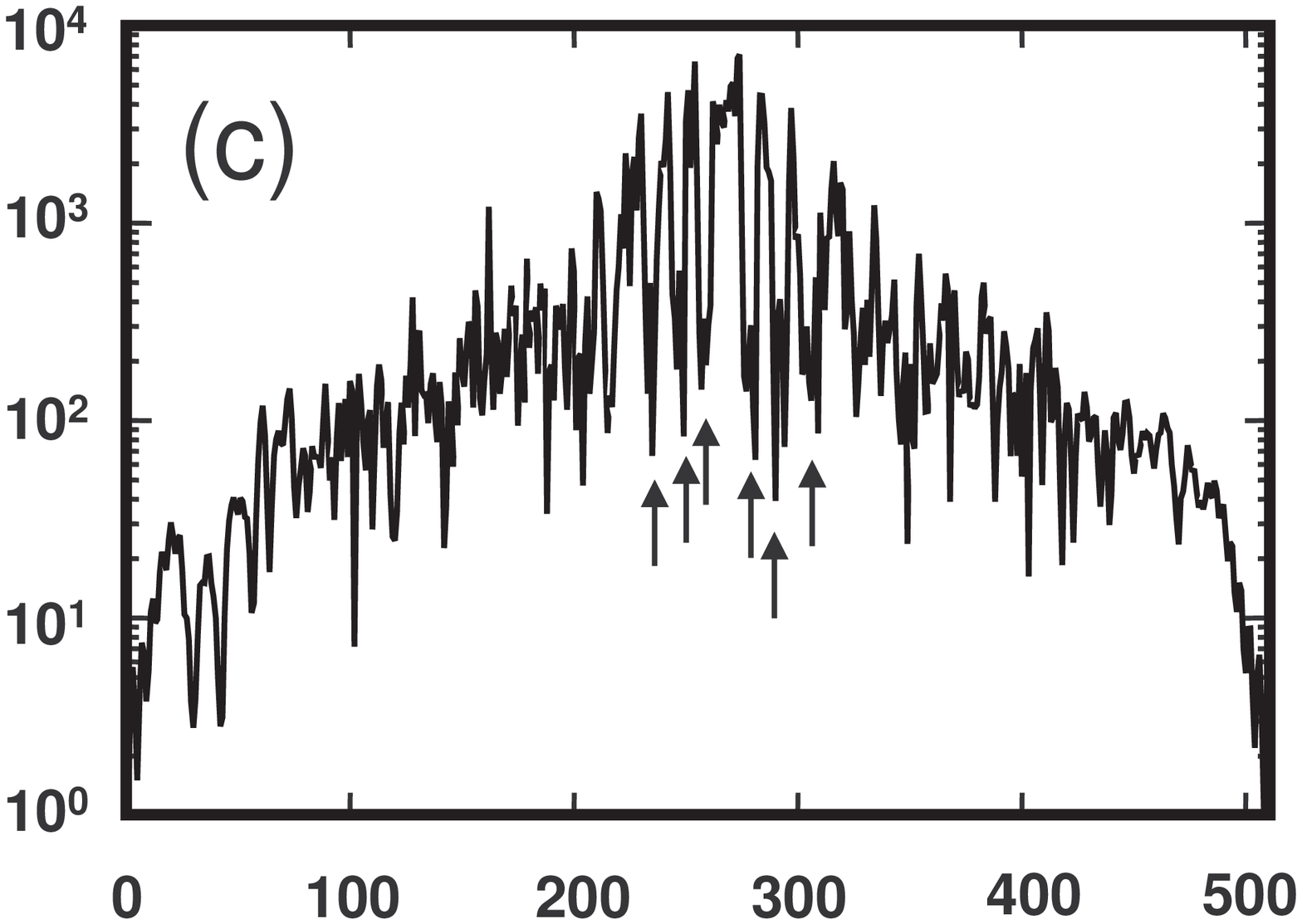}
\includegraphics[width = 4. cm,keepaspectratio=true]{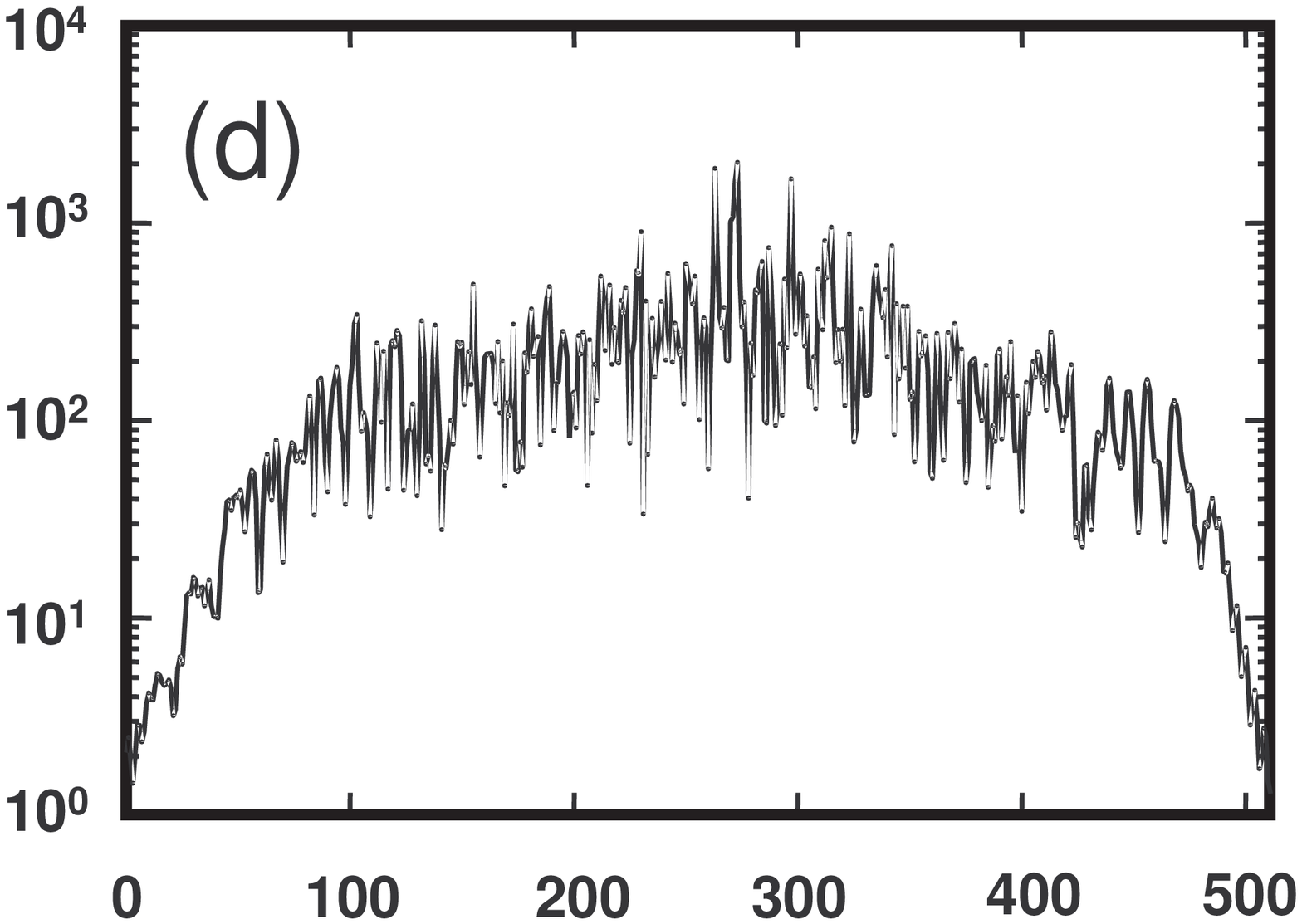}
\caption{Cuts of the 1-phase reconstructed images of the USAF
target in transmission with substraction of  the LO beam signal.
Images are obtained with k-space filtering with $\simeq 4.3\times
10^8$ (a), $8.7\times 10^6$ (b), $8.7 \times 10^4$ (c) and
$1.2\times 10^4$ (d) photo electron respectively for the sequence
of 3 images.} \label{fig_cut_usaf_14ph_f}
\end{center}
\end{figure}

Performing a double filtering, in space and in time, is an efficient
way to filter off the LO beam noise contributions. Quite good
signal-to-noise levels can be obtained by the time domain filtering
process which consist in subtracting the LO beam signal from the
1-phase hologram in order to cancel the $|{\cal E}_{LO}|^2$ term. To
illustrate this statement, we have calculated, from the sequence of
12 images, an approximation $I_{LO}$ of the image that should be
obtained with the LO beam alone. Since the phase of the hologram
interference pattern is shifted by $\pi/2$ from one CCD image to the
next, the CCD signal $I_{LO}$  is:

\begin{equation}\label{equ_I_LO}
    I_{LO}(x,y)= \frac{1}{12}\sum_{m=0}^{11}  I_m(x,y)
\end{equation}
It is then possible to remove the LO beam signal on each CCD image.
The corrected 1-phase  hologram $H"$ is then :
\begin{equation}\label{equ_holo_14phi}
    H"(x,y,0) =I_0(x,y) + I_4(x,y) +I_8(x,y)- 3I_{LO}(x,y)
\end{equation}
\begin{eqnarray}\label{equ_holo_14_bis}
\nonumber H"= \left(\sum_{m = 0,4,8}+ \frac{3}{12}\sum_{m = 0} ^{11}\right)  \left(  \left| {\cal E}\right|^2 + \left| {\cal E}_{LO}\right|^2 \right)\\
 ~~~~ + \sum_{m = 0,4,8} e^{+\omega_{CCD}t_m/4}~ {\cal E} {\cal E}_{LO}^* +  c.c.
\end{eqnarray}
As seen, the $\left| {\cal E}_{LO}\right|^2$ zero order term cancels
in $H"$.  From $H"$, we have reconstructed  LO-corrected 1-phase
images and cut, which are shown on Fig.\ref{fig_usaf_14ph_f} and
Fig.\ref{fig_cut_usaf_14ph_f}. The attenuation level (and thus the
total signal in photo electron units) is the same than for the
1-phase images and cuts of Fig.\ref{fig_usaf_1ph} and
Fig.\ref{fig_cut_usaf_1ph}. Here, the images and cuts quality is
much better than without LO correction in Fig.\ref{fig_usaf_1ph} and
Fig.\ref{fig_cut_usaf_1ph}. It is very close to the one obtained in
the 4-phase configuration in Fig.\ref{fig_usaf_4ph} and
Fig.\ref{fig_cut_usaf_4ph}. With respect to the 4-phase case, images
and cuts are slightly noisier here simply because the signal is
smaller by a factor 4, since the images are reconstructed from 4
times less CCD frames.

Note that these LO-corrected 1-phase results show that the phase
accuracy in 4-phase holography is not essential. The phase accuracy
has an effect on the weight of the twin image alias \cite{atlan_07},
but no significant effect on the weight of the zero order alias
itself. To cancel the zero order alias, it is sufficient to build
the hologram $H'$ with difference of images such a way the $|{\cal
E}|^2$ term is zero. This is ever realized whatever the
image-to-image phase shift is, since $\sum_{i=0}^{11} (-j)^m=0$ (see
Eq.\ref{equ_holo_4phi_bis}).

%
%
%

\section{Conclusion}

By comparing the reconstructed images obtained with holographic data
measured in off-axis and phase-shifting configuration, we have shown
that it is essential, for getting images with high SNR at low
illumination levels to fully filter-off the LO beam. In off-axis
holography, the LO beam yields the zero order image. It can be
filtered off by numerical removal of its contribution in k-space
domain (spatial filtering). This configuration also allows the
removal of the complex conjugate image. Nevertheless, this technique
is not efficient enough to reach optimal sensitivity. This is
particularly true when the camera exhibits image acquisition
defaults. These defaults can be visualized during data analysis (see
discussion about parasitic signal (arrow 2) visible on
Fig.\ref{fig_ima_1ph_700}(d)). In phase-shifting holography, the LO
beam is filtered-off by making images differences. The LO beam
component, which is the same on all the images, vanishes with images
differences. Although more efficient, phase-shifting is not
sufficient to fully cancel the LO beam contribution and its noise.
Single filtering (in space or time) is thus not enough, and double
filtering (in space and time) is needed if optimal sensitivity is
required. The key point of the time filtering process is to cancel
the $|{\cal E}_{LO}|^2$ term in the equation yielding the hologram.
We have to note that double filtering makes the measurement less
prone to camera technical noise and parasites. The comparison of the
images
represented throughout this article, which are calculated from the
same original data, is illustrative. Note that the quality of the
phase shift in multiple-phase, off-axis holography is not essential,
since quite as good results can be obtained with a single phase
hologram by subtracting roughly the LO beam contribution.

The authors acknowledge support from the French National Research
Agency (ANR) and the Centre de comp\'etence NanoSciences \^Ile de
France (C'nano IdF).



\end{document}